\documentclass[aps,prl,11pt,reprint,superscriptaddress,nofootinbib]{revtex4-1} 
\usepackage{amsmath,amssymb,amsthm,mathrsfs,bbm}
\usepackage{latexsym,amscd,amsbsy,amsfonts,dsfont}
\usepackage{graphicx}
\usepackage[utf8]{inputenc}
\usepackage{marvosym}
\usepackage{float}
\usepackage{slashed}
\usepackage{cancel}
\usepackage{multirow}
\usepackage{soul}
\usepackage{physics}
\usepackage{comment}
\usepackage{subcaption}
\usepackage[usenames,dvipsnames]{xcolor}
\usepackage{ulem}

\usepackage{todonotes}
\usepackage[
      colorlinks=true,
      linktocpage=true,
      breaklinks=true,
      linkcolor=Aquamarine,
      urlcolor=Purple,
      filecolor=black,
      citecolor=Purple,
      menucolor=black,
      pdfstartview=FitV,
      bookmarksopen=true
      ]{hyperref}
\usepackage{cleveref}
\definecolor{emerald}{rgb}{0.31, 0.78, 0.47}

\begin{document}
%--------------------------------------------------------------------

\title{Hawking radiation from an analogue bouncing geometry }
\author{Alberto Garc\'{i}a Mart\'{i}n-Caro}
\email{alberto.martin-caro@usc.es}
\affiliation{Departamento de F\'isica de Part\'iculas, Universidad de
Santiago de Compostela and \\
Instituto Galego de F\'isica de Altas Enerxias (IGFAE), E-15782
Santiago de Compostela, Spain}
\author{Gerardo Garc\'ia-Moreno}
\email{ggarcia@iaa.es}
\affiliation{Instituto de Astrof\'{\i}sica de Andaluc\'{\i}a (IAA-CSIC), Glorieta de la Astronom\'{\i}a, 18008 Granada, Spain}
\author{Javier Olmedo}
\email{javolmedo@ugr.es}
\affiliation{Departamento de F\'{\i}sica Te\'orica y del Cosmos, Universidad de Granada, Granada-18071, Spain}
\author{Jose M. S\'anchez Vel\'azquez}
\affiliation{Instituto de F\'isica Te\'orica UAM/CSIC, c/ Nicol\'as Cabrera 13-15, Cantoblanco, 28049, Madrid, Spain}
\email{jm.sanchez.velazquez@csic.es}

%------------------------------------------------
\begin{abstract}
We propose a setting that simulates Hawking radiation from an analogue bouncing geometry, i.e., a collapsing geometry that reverts its collapse after a finite time, in a setup consisting of a coplanar waveguide terminated in superconducting quantum-interference devices at both ends. We demonstrate experimental feasibility of the proposed setup within the current technology. Our analysis illustrates the resilience of Hawking radiation under changes in the physics at energy scales much larger than the temperature, supporting the idea that regular alternatives to black holes would also emit Hawking radiation. 
\end{abstract}
%------------------------------------------------

\maketitle
%------------------------------------------------
\paragraph*{\textbf{Introduction.}} 
%------------------------------------------------

Hawking radiation –the spontaneous creation of particles in a collapsing geometry (close to) forming a black hole– is one of the most important processes in Quantum Field Theory in Curved Spacetimes (QFTCS)~\cite{Birrell1982,Wald1995}. This phenomenon is at the heart of one of the most important open problems in fundamental physics: whether quantum gravity is intrinsically non-unitary~\cite{Hawking1976,Unruh2017,Almheiri2012}. In the past, it was believed that Hawking radiation was tightly related to the presence of a causal horizon. However, more recent analyses have identified the exponential peeling of geodesic curves within these kind of geometries to be the fundamental ingredient behind Hawking radiation~\cite{Visser2001,Barcelo2006,Barcelo2006b}. Actually, it is precisely this connection what has lead to many proposals to reproduce Hawking radiation experimentally within the so-called analogue gravity program, see~\cite{Barcelo2005} for a review. Three of the most promising analogue systems~\cite{Barcelo2018} are Bose-Einstein condensates~\cite{Garay1999,Garay2000,Barcelo2001}, with a recent experimental verification~\cite{Steinhauer2015,Nova:2019,Kolobov2019}, although the spontaneous origin of the radiation is still a topic of debate within the community~\cite{IsoardThesis,Leonhardt2016,Steinhauer2018}; surface waves in water flows~\cite{Schutzhold2002}, which have been implemented experimentally~\cite{Rousseaux2007,Weinfurtner2010}, although there is still debate about the interpretation of the result as stimulated Hawking radiation and the nature of the horizon~\cite{Weinfurtner2013,Michel2014,Coutant2016,Euve2014}, see also the more recent results~\cite{Euve2014,Euve2018}; and also nonlinear optics systems, where we can highlight the use of electromagnetic waveguides~\cite{Schutzhold:2004tv}, optical fibres~\cite{Philbin:2007ji,Belgiorno2010,Rubino2011} (see also~\cite{Drori2018} for more recent work), and quantum fluids of light obeying an effective Gross-Pitaevskii equation~\cite{Carusotto2004} that have been engineered in the laboratory~\cite{Nguyen2015}. 

There are two basic ingredients required for the Hawking effect:  i) a locally Lorentz invariant quantum field theory propagating on top of a curved geometry, ii) having a geometry displaying an exponential peeling of the geodesics (i.e. on the verge of forming a causal horizon). Any modification of these two ingredients is a test of the robustness of the Hawking radiation. Whereas modifications of the former have been widely explored in analogue systems before, see~\cite{Barcelo2018} for a careful discussion and a guide to the literature, modifications of the latter like the ones that we consider in our article have not been explored so much and our system is specially well-suited for it. In fact, from the quantum gravity side it is specially interesting to design analogue systems of bouncing geometries as alternatives to singular black hole collapse~\cite{Stephens1993,Barcelo2014,Husain2021,Husain2022,Kelly2020}, since they have not been explored that much in analogue systems in the past.

Quantum information techniques and devices currently play a very important role in QFTCS and in quantum gravity. In many cases, these techniques allow to understand processes in QFTCS as simple quantum circuits, allowing to simplify computations and clarify the interpretation of the results \cite{Agullo2021,Brady:2022ffk}. Furthermore, many of the systems used for quantum simulation and computation purposes can also be used as analogue systems. In particular, superconducting quantum interference devices (SQUIDs) are among the most popular systems that are used nowadays as potential quantum simulators \cite{Sab_n_2016,Terrones_2021}.

In this Letter, we propose a way to simulate Hawking radiation in a (1+1)-dimensional system consisting of a coplanar waveguide (CPW) terminated in two SQUIDs at both of its endings. As pointed out in \cite{Johansson:2009zz}, by tuning the magnetic flux that threads the SQUIDs, one can tune the boundary conditions of the real free scalar field encoding the phase of the electric field propagating in the waveguide. In particular, we can make those boundary conditions to follow uniformly accelerating trajectories, a situation that has been widely studied in the literature (specially for a single boundary) since it gives rise to Hawking-like radiation and the formation of the analogue of a causal horizon \cite{Birrell1982}. We study a variation of these trajectories in which the field is confined between two boundaries which start at rest, accelerate, then decelerate, and stop. Such trajectories correspond to a system which is on the verge of forming a causal horizon but stops at some point, giving rise to a regular causal behaviour during the whole process. This can be interpreted as an analogue to a collapsing geometry that at some point reverts the collapse through a bounce and gets stabilized in a regular configuration.

The main lesson that we draw up from this work is that, with state of the art technologies, there are {\it experimentally achievable} trajectories of the boundaries representing effective bouncing geometries where there is nearly thermal particle production for long wavelength modes, up to effects associated with the boundary accelerating only for a finite time, which translate into oscillations in the frequency bands \cite{Good:2013lca,Good:2019tnf}. We work in natural units $\hbar = c =1$. 

%------------------------------------------------
\paragraph*{\textbf{CPW+SQUIDs setup}}
%------------------------------------------------ 

We consider a superconducting coplanar waveguide (CPW) terminated in a SQUID at both ends. The electromagnetic field at each point in the waveguide and at a given time can be described by a real scalar field $\phi(t,x)$, the so-called phase field, which is obtained as the time-integral of the electric field \cite{Johansson:2009zz}. The effective equation of motion for this field inside the cavity is the massless Klein-Gordon equation with effective propagation speed $ v = 1 /\sqrt{\mathcal{L} \mathcal{C}}$, being $\mathcal{L}$ and $\mathcal{C}$ the characteristic inductance and capacitance of the CPW per unit length, respectively. 

Consider now that we apply an external, time-dependent magnetic flux through each of the SQUIDs independently. We work in the regime in which the plasma frequency of the SQUIDs, $\omega_p$, is much larger than the characteristic driving frequency $\omega_d$ of the applied fluxes,
which we represent as $\Phi_L(t)$ and $\Phi_R(t)$, corresponding to the left and right endings, respectively. Under these conditions, the scalar field in the CPW satisfies the following Robin boundary conditions \cite{Johansson2010,Bosco:2019ayk}:
%------------------------------------------------
\begin{align}
    & \phi(t,0)-\delta L(\Phi_L)\partial_x\phi(t,x)\!\left.\right|_{x=0} = 0,  \\
    & \phi(t,L)-\delta L(\Phi_R)\partial_x\phi(t,x)\!\left.\right|_{x=L} = 0.
\end{align}
%------------------------------------------------
$L$ represents the length of the CPW, with $\delta L(\Phi)=\Phi_0/2\pi\qty(2L_0I_c\abs{\cos(\pi\Phi/\Phi_0)})^{-1}$ where $\Phi_0 = \pi \hbar/e$ is the magnetic flux quantum and $I_c$ the SQUID's critical current \cite{Johansson2010}. Now we focus on excitations whose wavelength is much greater than the displacement $\delta L \ll \lambda$ and the regime $\delta L \ll L_0$. In this regime, the Robin boundary conditions reduce  approximately to Dirichlet boundary conditions, i.e. $ \phi(t,-\delta L(\Phi_L)) = \phi(t,L_0-\delta L(\Phi_R)) = 0$, \cite{Johansson:2009zz, Johansson2010}. 

These boundary conditions model a cavity enclosed by a pair of perfectly reflecting mirrors at positions $ f(t)=-\delta L(\Phi_L)$ and $g(t)=L_0-\delta L(\Phi_R)$. Therefore, by dynamically tuning the fluxes, the CPW$+$SQUIDs system simulates a cavity with moving Dirichlet boundaries; as long as the conditions $\omega_d \ll \omega_p$,  $\lambda \gg \delta L$  and $\delta L \ll L_0$ are fulfilled. Indeed, a SQUID-terminated CPW setup (like the one we propose here) has already been used in the experimental verification of the dynamical Casimir effect~\cite{Johansson2011nat}. 

%------------------------------------------------
\paragraph*{\textbf{Hawking radiation in SQUIDs.}} 
%------------------------------------------------

Up to now we have argued that the system that we are considering is amenable to simulate a QFT of a real scalar field confined within two moving boundaries at which we impose perfectly reflecting boundary conditions. This means that the field identically vanishes there. Moreover, it is well-known that a single moving mirror in (1+1) Minkowski spacetime following a Rindler-like trajectory is able to generate  thermal radiation and it can be understood as an analogue of Hawking radiation~\cite{Candelas1976,Davies1976,Candelas1977,Davies1977,Birrell1982}. It is produced in the left-moving modes if the mirror accelerates to the right and vice-versa, and, due to the eternal motion of the boundary, an infinite number of particles is produced. 
Obviously, we cannot simulate an eternal acceleration. However, Hawking radiation does not necessarily require it, as we have discussed in the introduction. In fact, in~\cite{Good:2019tnf} a trajectory in which the boundary moves within a finite time was proposed. It gives rise to a finite number of particle creation displaying an almost thermal spectrum.

\begin{figure}
\begin{center}
\includegraphics[width=0.8\linewidth,height=0.27\textheight]{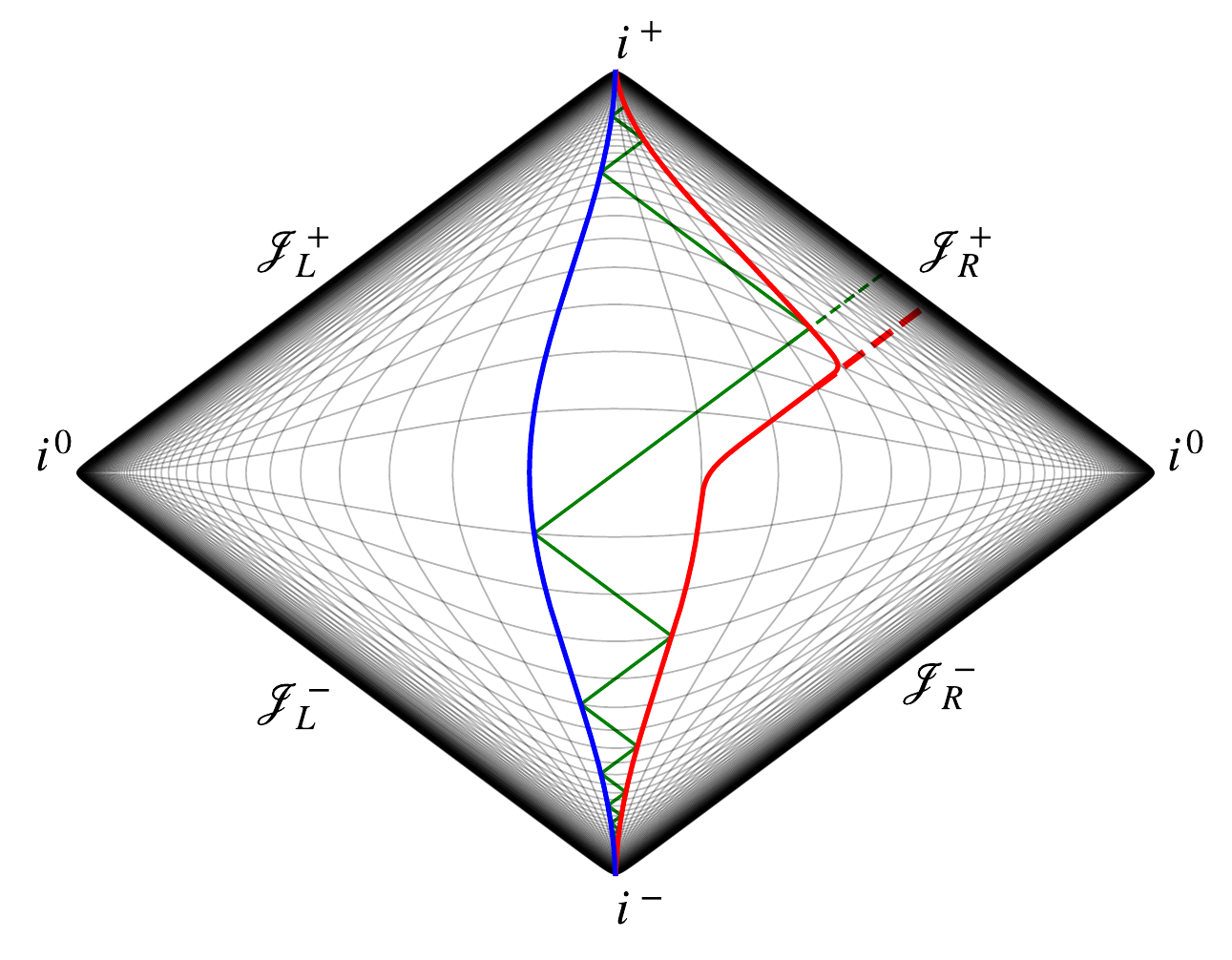}
\caption{Example of spacetime trajectories of the left (blue-solid) and right (red-solid)  mirrors. We show a configuration where the right mirror  accelerates for a finite time. We also include the trajectory if it were to undergo an indefinite acceleration (red-dashed). In green we represent a light ray propagating in the cavity (solid and dashed if the right mirror accelerates for a finite or infinite time interval, respectively).}
\label{fig:Penrose}
\end{center}
\end{figure}

Here, we will focus on a different kind of trajectories that are specially interesting from the perspective of its implementation in the CPW+SQUID setup described in the previous section. Here the quantum field is confined between two (not only one) perfectly reflecting boundaries for the frequencies of interest. Our aim here is twofold: i) to determine under which conditions is the radiation spectrum still (nearly) thermal, and ii) to analyze whether it is possible to simulate these trajectories in the CPW+SQUIDs system with current technology. We note that in the case in which one has only one boundary, or one of the two boundaries accelerates indefinitely, the thermal flow reaches the region of $\mathscr{J}^+$ supported in the causal future of the nontrivial accelerating part of the boundary trajectory (see \cref{fig:Penrose}), so it can be easily identified.
But with two boundaries and finite acceleration times, the situation is more complicated. Even if one of them does not move, all the produced radiation will be confined within the cavity at late times when the motion of the other boundary has stopped. Hence, thermal Hawking-like particles will be superposed to particle creation due to the transients of the trajectories of the mirrors, i.e. the parts in which the acceleration changes significantly, and also there will be a part of the particle creation which will be stimulated (not from the vacuum state). 

We suggest two ways of bypassing this problem, although we show that the second one seems to be much more feasible within the current experimental capabilities. The first of these ways is to choose a sufficiently adiabatic transition between the constant acceleration phases. In this way, the particle production by the transients is minimized but these trajectories require $\delta L\gg L$, which is not experimentally realizable nowadays.  The second consists in making the transients sharp, so that
the energy scales associated with the modes excited during the almost uniformly accelerated phase and the transient regimes are parametrically separated. In that way, although the computed spectrum will deviate from thermality for ultraviolet modes, long wavelength modes will display a nearly thermal spectrum, up to oscillations due to the finite time interval of acceleration and (order unity) grey body factors. This reaffirms the resilience of long-wavelength Hawking radiation to modifications of the physics at very short length scales, as it has been already suggested in the literature through the analysis of a wide variety of models~\cite{Unruh1994,Corley1996,Corley1996b,Corley1997,Corley1998,Unruh2004,Barcelo2008,Finazzi2012,Robertson2015}. 

We need then to study the dynamics of a massless, Klein-Gordon scalar field $\phi(t,x)$ inside a one-dimensional cavity propagating at the speed of light, and with the Dirichlet boundary conditions $\phi(t,x=f(t))=\phi(t,x=g(t))=0$, with $f(t),g(t)$ the trajectories of the left and right mirrors, respectively. We can decompose the field in Fourier modes
\begin{equation}\label{eq:fourier}
 \phi(\xi)=  \sum_{n=1}^{\infty} \phi_n(t)\sin (n \pi \xi(t)),
\end{equation}
with $\xi=L_0\frac{x-f(t)}{g(t)-f(t)}$ (in the following we set $L_0=1$ unless otherwise specified). It is worth noting that $\xi$ defines a new coordinated system where the field satisfies static boundary conditions  $\phi(t,\xi=0)=\phi(t,\xi=1)$, and where the transformed metric takes the form of an acoustic metric \cite{prd1}. 

The Fourier coefficients $\phi_n(t)$ satisfy the equations of motion
\begin{equation}
    \ddot{u}_n+ \sum_{m=1}^\infty R_{nm}\dot{u}_m +\sum_{m=1}^{\infty} S_{nm}u_m=0,
    \label{shortEoM}
\end{equation}
where, if we define $L(t)=g(t)-f(t)$,  \begin{equation}\label{eq:RnmSnm-b}
    \begin{split}
          &S_{mn}=\delta_{mn}\bigg[\left(\frac{n\pi}{L}\right)^2\left(1-\dot f^2-\dot f\dot L-\frac{\dot L^2}{3}\right)\\
          & +\frac{\dot L^2}{2L^2}\left(1-\frac{2}{n\pi}\right)+\frac{\ddot L}{2 n\pi L}\bigg]+ ((-1)^{m+n}-1)\\
          & \times\bigg[\frac{2[\ddot fL+\ddot LL-2\dot f\dot L-2\dot L^2]m}{(m^2-n^2)\pi L^2}
          -\frac{8[\dot f\dot L+\dot L^2]mn^3}{(m^2-n^2)^2L^2}\bigg], \\
          &R_{mn}= -\delta_{mn}\frac{\dot L }{L}-(1-\delta_{mn})\frac{4[(-1)^{m+n}(\dot f+\dot L)-\dot f]mn}{(m^2-n^2)L}.
    \end{split}
\end{equation}
We solve this set of equations numerically with an explicit embedded Prince-Dormand-Runge-Kutta (8,9) method. In order to do so, we truncate the maximum number of modes $N$, concretely, to the values $N=128,\,256,\,512$, and adopt a Richardson extrapolation to obtain the limit $N\to\infty$. We use this method to construct the $in$-basis of complex, positive frequency solutions ${\bf u}^{(J)}(t)=\big(u_{1}^{(J)}(t),u_{2}^{(J)}(t),\ldots\big)$, and its complex conjugate, for $J=1,2,\ldots$ This basis is well adapted to the natural $in$ vacuum state at early times, where the boundaries remain static. It is normalized with respect to the Klein-Gordon product. The quantum field can be written as 
\begin{equation}\label{eq:mode-aadagg}
\hat\phi_n(t) =  \sum_{J=1}^\infty  u_{n}^{(J)}(t)\hat a_J+\bar u_{n}^{(J)}(t)\hat a_J^{\dagger},
\end{equation}
where $(\hat a_J,\hat a_{J'}^{\dagger})$ are the annihilation and creation variables. Particle production at late times will be given by the coefficients 
\begin{align}\label{eq:beta}
\beta_{IJ} = -\langle \bar {\bf u}^{(J)}(t), {\bf w}^{(I)}(t)\rangle.
\end{align}
of the Bogoliubov transformation between the $in$ and $out$ states basis, the latter determined by the basis of complex solutions $\big({\bf w}^{(I)}(t), \bar {\bf w}^{(I')}(t)\big)$.

Among several configurations for the trajectories of the boundaries that we have studied, we show here a configuration in which the boundaries are initially static, then only the right boundary accelerates, approaching the effective speed of light, and it finally decelerates until it stops completely, in a time-symmetric way. The explicit trajectories for the boundaries are $f(t)= 0$ and
\begin{align}\nonumber
   %f(t)= \frac{s}{4\kappa}+\frac{1}{4\kappa}[\log(\cosh(\kappa(t-t_0)))-\notag\\
   %-&\log(\cosh(s-\kappa(t-t_0)))],\\
   g(t)= & 1 +\frac{s}{2\kappa}+\frac{1}{2\kappa}\bigg[\log\Big(\cosh\big(\kappa(t-t_0)\big)\Big)-\notag\\
   -&\log\Big(\cosh\big(s-\kappa(t-t_0)\big)\Big)\bigg],
   \label{eq:Traj}
\end{align}

In this configuration, at $t \ll t_0$, the right mirror is nearly static at initial position $x_{in}=1$, while at $t\gg T\simeq t_0+\epsilon$ (with $\epsilon=s/\kappa$) its final position $x_{out} =(1+\epsilon)$. In the interval $[t_0,t_0+\epsilon]$ the right mirror accelerates close to the speed of light. We depict this configuration in the Penrose diagram of Fig. \ref{fig:Penrose}.

During the two phases of almost constant acceleration, we expect a thermal flux of produced particles due to
the exponential delay in affine parameter of null rays. Actually, the same must happen in the phase of constant deceleration, since the peeling is still exponential (although with different sign, but this is irrelevant for particle creation). 
Therefore, this trajectory would be mimicking light rays propagation in a bouncing geometry from a black hole to a white hole transition, as the one suggested in~\cite{Barcelo2014}. 
However, we also expect that the particle production will be almost, but not exactly, thermal in both the accelerating and decelerating phase. This additional particle production comes from the transients during acceleration from and deceleration to the resting positions of the boundaries. 
However, it is in the ultraviolet modes where particle creation from the transients will be produced and where deviation from thermality should be observed (we note that the overproduction in these modes due to the transients will be superpose to the  thermal particle production).

\begin{figure}
    \centering
    \includegraphics[width=8.6cm]{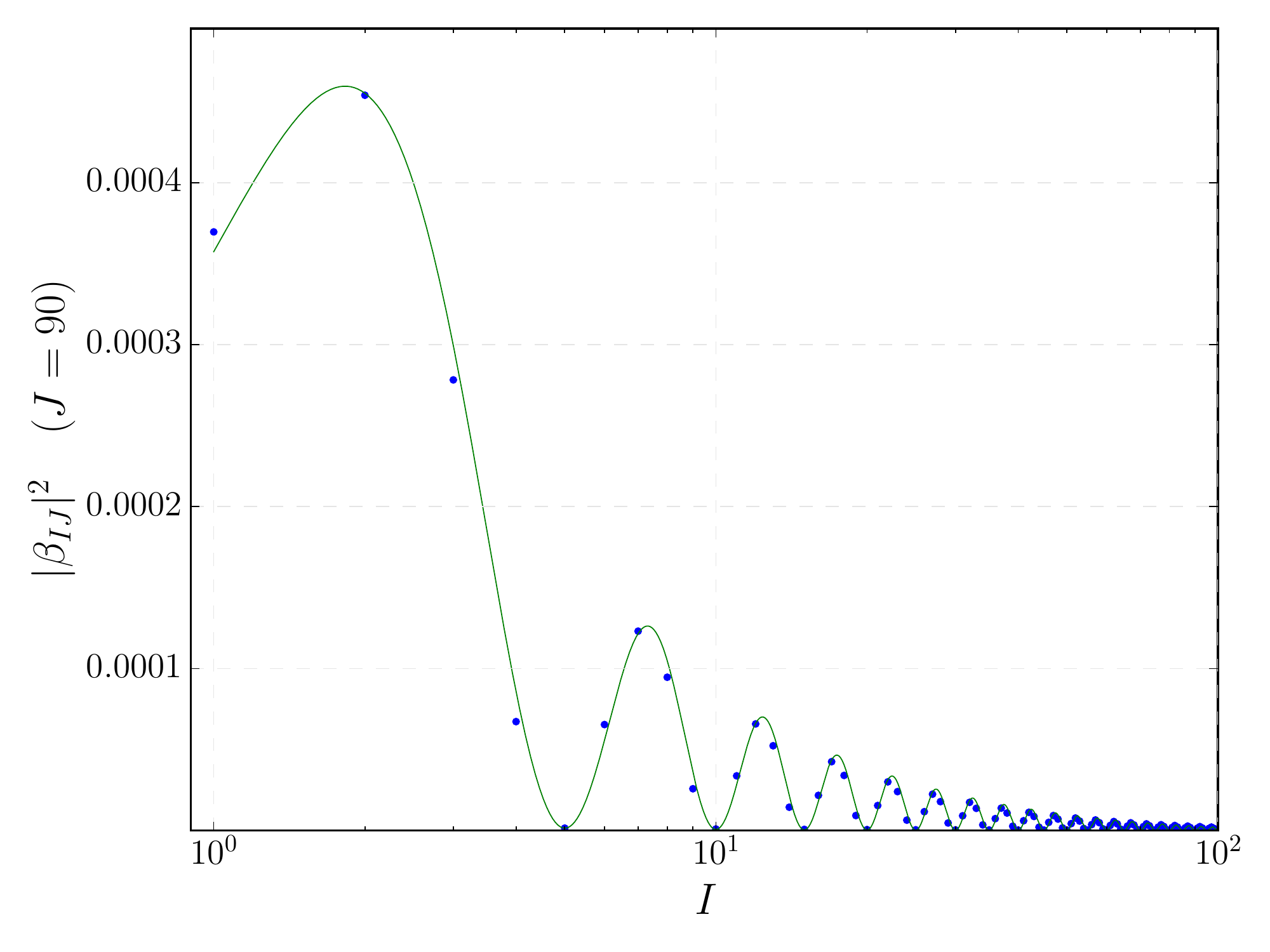}\\
    \includegraphics[width=8.6cm]{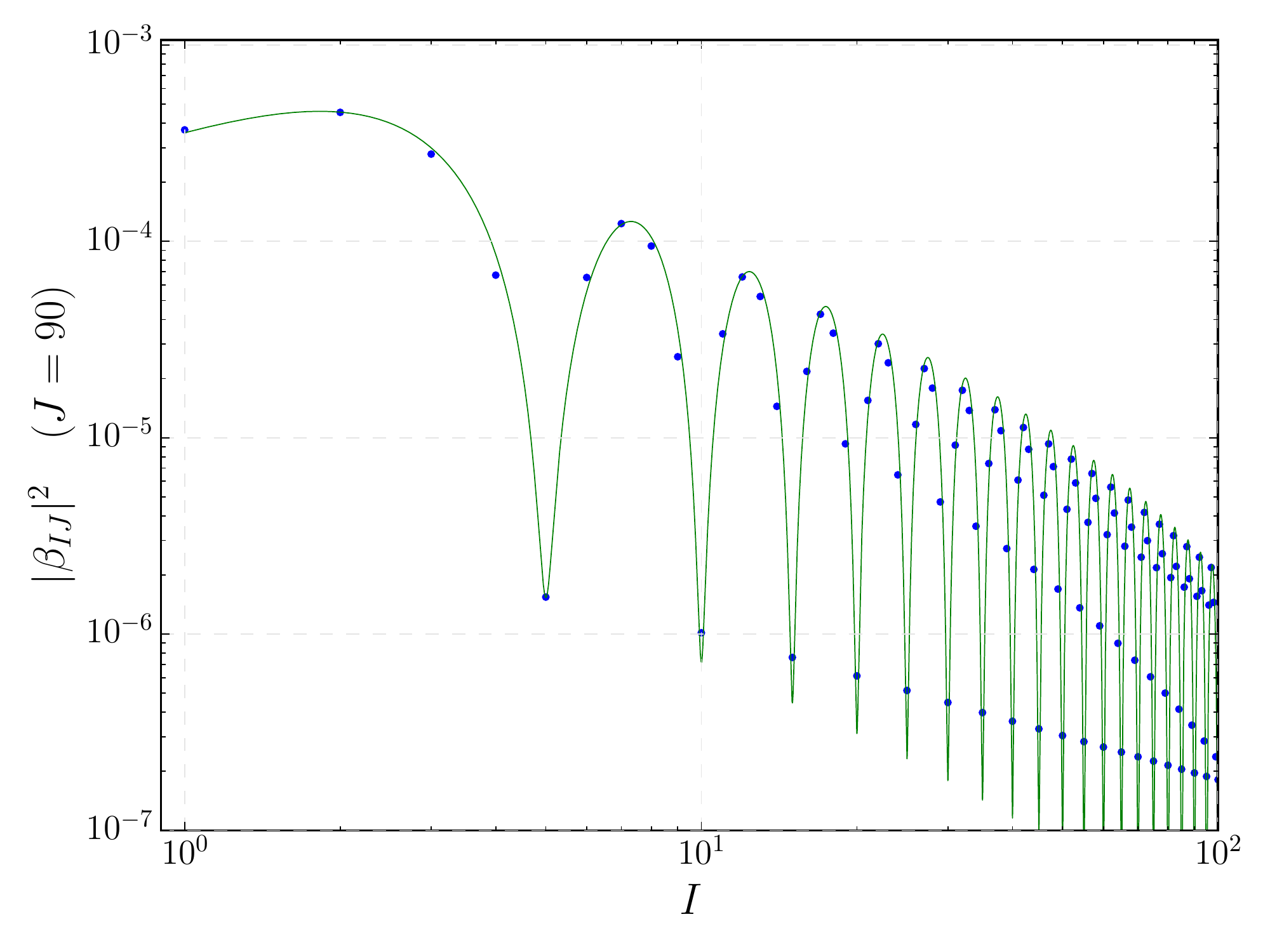}
    \caption{This is the spectrum for a  trajectory of the left boundary determined by $s=150$ and $\kappa=600$. Here, we show it for $J = 90$. The continuous lines represent the Planckian distribution in Eq. \eqref{eq:beta-adj} with 
    $A(\kappa)=7.3\cdot 10^{-3}$ and $B(\kappa)=0.91$. The dots represent the spectrum obtained in our simulations. Upper panel: The $y$ axis is in linear scale. Lower panel: The $y$ axis is in logarithmic scale.}
    \label{fig:betas}
\end{figure}

To confirm the previous statements, we plot the beta-coefficients in Eq. \eqref{eq:beta} in Fig. \ref{fig:betas}. For this trajectory $\kappa=600$ and $s=150$. This corresponds to a final static configuration for the left and right boundaries determined by $\epsilon\simeq 0.25$. This final position of the boundary lies within the experimental values considered in Ref. \cite{Bosco:2019ayk}. This simulation is in very good agreement with a spectrum of the form
\begin{equation}\label{eq:beta-adj}
|\beta^{\text{(fit)}}_{IJ}|^2 = \frac{2 \Delta \omega_I\Delta \omega_J}{\pi\kappa\omega_J}\frac{\Gamma(\kappa,\epsilon)}{(e^{2\pi\omega_I/\kappa}-1)},
\end{equation}
where $\Delta\omega_I=\pi/ (L_0+\epsilon)$, $\Delta\omega_J=\pi/ L_0$   $\big($in the continuum   $\Delta\omega_I\Delta\omega_J\to d\omega d\omega'$$\big)$, $\omega_I= \pi I/ (L_0+\epsilon)$, $\omega_J= \pi J/ L_0$, and
\begin{equation}\label{eq:Gamma}
\Gamma(\kappa,\epsilon)\simeq\left[A(\kappa,\epsilon)+B\sin^2\left(\epsilon \, \omega_I\right)\right],
\end{equation}
is a grey body factor that accounts for the finiteness of the duration of the acceleration, producing the oscillations described in very good approximation by a sinusoidal function with a well-defined frequency characterized in Eq. \eqref{eq:Gamma}. It has a simple dependence on $\epsilon$. In the simulations we have studied, $A(\kappa,\epsilon)$ has some dependence on $\kappa$ and $\epsilon$, with its value being around $10^{-2}$, while $B \simeq 0.9$, regardless of $\kappa$ and $\epsilon$. We also notice that the difference in the normalization between \eqref{eq:beta-adj} and Eq. (4.61) of Ref. \cite{Birrell1982} (obeying $\Gamma$) is due to the presence of two boundaries, instead of only one. We have studied other simulations for $\epsilon\in [0.125,0.5]$ and $\kappa\in [50,1200]$, and the behavior is qualitatively similar for all those trajectories, for all frequencies $\omega_I$ with $I\in[1,100]$ and $J\in[30,200]$ (frequencies outside these intervals deviate from thermality). 
%

%------------------------------------------------
\paragraph*{\textbf{Experimental setting.}} 
%------------------------------------------------

Regarding the experimental implementation of our proposal, we must note that the speed of propagation $v$ of the phase field $\phi$ and the cavity lengths $L$, and its effective change $\delta L$ due to the SQUIDs must satisfy $v\ll \omega_p \delta L<\omega_p L$ at all times. In this regime, the boundaries induced by the SQUIDs can accelerate and reach $v$ asymptotically. In \cite{Johansson2010} the plasma frequency is $\omega_p = 37.3$~GHz and $\delta L/L_0\simeq 0.25$. If we assume a reasonable value for the length of the CPW of $L_0=1.0$~cm, our trajectories can be simulated on actual cavities. However, for a typical speed \cite{Johansson2010} of the phase field $v\simeq 10^8$~(m/s), the maximum Hawking temperature one could reach is about $T_H\simeq 0.2$~K. This is below the experimental noise temperature reported in \cite{Johansson2011nat}. In order to reach higher temperatures, one should consider characteristic inductance and capacitance parameters of the CPW such that the speed $v$ of the phase field is smaller by one or two orders of magnitude. Since these quantities are related by $v=1/\sqrt{\mathcal{L}\mathcal{C}}$, one must consider  CPWs with modified inductances and capacitances, either by suitable changes in their geometry or by considering substrates with large dielectric permittivities \cite{chen1997,goppl2008} (there exist materials like perovskite or polymers with permittivities higher than $10^4$), reducing so the speed $v$ of propagation of the phase field of the electric field on the CPW. Reducing $v$ is in agreement with the condition  $v\ll \omega_p \delta L \simeq 1.2\cdot 10^{8}$~(m/s). For instance, for values of $v\simeq 10^6$~(m/s) and a maximum acceleration of the boundary $\kappa\simeq \omega_p^2 \delta L \simeq  3.5\cdot 10^{18}$~(m/s${}^2$), the maximum Hawking temperature can be as large as $T\simeq 30$~K, well above the system noise temperature reported in \cite{Johansson2011nat}. It is illustrative to note that the simulation reported in Fig. \ref{fig:betas} corresponds to an acceleration $\kappa\simeq 6\cdot 10^{16}$~(m/s${}^2$) of the boundary. For $v\simeq 10^6$~(m/s), the corresponding Hawking temperature will be $T_H\simeq 0.5$~K.

%------------------------------------------------
\paragraph*{\textbf{Conclusions.}} 
%------------------------------------------------

We have presented numerical evidence of Hawking-like radiation in a setup involving a CPW with two SQUIDs attached to its endings. The phase of the electromagnetic field propagating along the CPW can be well-described by a real, scalar quantum field $\phi(t,x)$. The boundary conditions of the system are determined completely in terms of the magnetic fluxes that thread the SQUIDs at both endings and, in principle, they can be tuned at will as functions of time. For sufficiently long wavelengths, the boundary conditions correspond to Dirichlet boundary conditions. In this way, we have a system whose excitations correspond to a quantum real scalar field confined within two moving boundaries. 

We have computed the spectrum of particles produced in different trajectories of the boundaries. Here, we show one configuration where the right  boundary approaches a phase of uniform acceleration. However, instead of allowing the boundary to accelerate forever, we make it stop at a given time. It is well-known that a thermal flow of particles will be produced during the uniformly accelerating part of the trajectory. However, on top of this thermal flow we also have effects due to finite time acceleration of the boundary and the ones associated with the transients. To avoid the mixing in some of the modes between the thermal and non-thermal production, we have engineered trajectories for which the changes in the acceleration are abrupt enough so that the production due to those transients mainly affects high energetic modes. In principle, the proposed trajectories can be simulated by the SQUIDs at the ends of the CPW. Furthermore, the boundary conditions become of Robin type for sufficiently short wavelengths, therefore we have focused on this long wavelength regime for which we confirm numerically the nearly thermal flow of particles. 

The trajectories we have considered  mimic the exponential peeling of geodesics in a (horizonless) bouncing geometry, which produces a nearly-thermal spectrum of created particles. 
This scenario has been proposed before as an alternative to the standard evaporation paradigm through Hawking radiation, and it is expected to regularize the singularity within the core of black holes. Our results suggest that these trajectories are amenable to simulation within the current technology and confirm the almost thermal particle production on the long wavelength modes. 

Finally, a comment regarding the quantum nature of the field is in order. Whenever the system is close enough to the vacuum, the excitations of the system do display a sharp quantum character and they can be interpreted within the QFT perspective as particle creation. Actually, there might be squeezed states for which this quantum character is even more highlighted~\cite{Parikh2020,Parikh2020b,Agullo2021}. However, notice that if the original state is a semiclassical state, in the sense of being a state whose expectation values of the quadratures are highly peaked, the evolution of the state is well-captured as a classical system. Hence, one of the key points to ensure a purely quantum behaviour the system is to ensure that the initial state is ``quantum enough" e.g. squeezed or vacuum states. A future interesting follow-up of this work would be to characterize for which states is the quantum nature of the phenomenon enhanced. Actually, it would be interesting to develop optimization techniques for this purpose.

In summary, our proposal may lead to an experimental setup that allows for the simulation of Hawking-like radiation in a simple and controlled fashion analogue system, paving the way towards the experimental verification of Hawking-like radiation in analogue bouncing geometries.

%------------------------------------------------
\paragraph*{\textbf{Acknowledgements.}}
%------------------------------------------------
The authors would like to thank Iv\'an Agull\'o, Carlos Barcel\'o and Luis Cort\'es Barbado for invaluable conversations. AGMC acknowledges financial support from the Spanish Ministry of Education, Culture, and Sports (Grant No. PID2020-119632GB-I00), the Xunta de Galicia (Grant No. INCITE09.296.035PR and Centro singular de investigación de Galicia accreditation 2019-2022), the Spanish Consolider-Ingenio 2010 Programme CPAN (CSD2007-00042), and the European Union ERDF. AGMC is also grateful to the Spanish Ministry of Science, Innovation and Universities, and the European Social Fund for the funding of his predoctoral research activity (Ayuda para contratos predoctorales para la formaci\'on de doctores 2019). GGM is funded by the Spanish Government fellowship FPU20/01684 and acknowledges financial support from the grant CEX2021-001131-S funded by MCIN/AEI/10.13039/501100011033. Financial support is provided by the Spanish Government through the projects PID2020-118159GB-C43, PID2019-
105943GB-I00 (with FEDER contribution). JO is supported by the ``Operative Program FEDER2014-2020 Junta de Andaluc\'ia-Consejer\'ia de Econom\'ia y Conocimiento'' under project E-FQM-262-UGR18 by Universidad de Granada. JMSV acknowledges the support of the Spanish Agencia Estatal de Investigaci\'on through the grant “IFT Centro de Excelencia Severo Ochoa CEX2020-001007-S".
%------------------------------------------------

%------------------------------------------------

\bibliography{dce_prl_biblio}

%merlin.mbs apsrev4-1.bst 2010-07-25 4.21a (PWD, AO, DPC) hacked
%Control: key (0)
%Control: author (72) initials jnrlst
%Control: editor formatted (1) identically to author
%Control: production of article title (-1) disabled
%Control: page (0) single
%Control: year (1) truncated
%Control: production of eprint (0) enabled
\begin{thebibliography}{67}%
\makeatletter
\providecommand \@ifxundefined [1]{%
 \@ifx{#1\undefined}
}%
\providecommand \@ifnum [1]{%
 \ifnum #1\expandafter \@firstoftwo
 \else \expandafter \@secondoftwo
 \fi
}%
\providecommand \@ifx [1]{%
 \ifx #1\expandafter \@firstoftwo
 \else \expandafter \@secondoftwo
 \fi
}%
\providecommand \natexlab [1]{#1}%
\providecommand \enquote  [1]{``#1''}%
\providecommand \bibnamefont  [1]{#1}%
\providecommand \bibfnamefont [1]{#1}%
\providecommand \citenamefont [1]{#1}%
\providecommand \href@noop [0]{\@secondoftwo}%
\providecommand \href [0]{\begingroup \@sanitize@url \@href}%
\providecommand \@href[1]{\@@startlink{#1}\@@href}%
\providecommand \@@href[1]{\endgroup#1\@@endlink}%
\providecommand \@sanitize@url [0]{\catcode `\\12\catcode `\$12\catcode
  `\&12\catcode `\#12\catcode `\^12\catcode `\_12\catcode `\%12\relax}%
\providecommand \@@startlink[1]{}%
\providecommand \@@endlink[0]{}%
\providecommand \url  [0]{\begingroup\@sanitize@url \@url }%
\providecommand \@url [1]{\endgroup\@href {#1}{\urlprefix }}%
\providecommand \urlprefix  [0]{URL }%
\providecommand \Eprint [0]{\href }%
\providecommand \doibase [0]{http://dx.doi.org/}%
\providecommand \selectlanguage [0]{\@gobble}%
\providecommand \bibinfo  [0]{\@secondoftwo}%
\providecommand \bibfield  [0]{\@secondoftwo}%
\providecommand \translation [1]{[#1]}%
\providecommand \BibitemOpen [0]{}%
\providecommand \bibitemStop [0]{}%
\providecommand \bibitemNoStop [0]{.\EOS\space}%
\providecommand \EOS [0]{\spacefactor3000\relax}%
\providecommand \BibitemShut  [1]{\csname bibitem#1\endcsname}%
\let\auto@bib@innerbib\@empty
%</preamble>
\bibitem [{\citenamefont {Birrell}\ and\ \citenamefont
  {Davies}(1984)}]{Birrell1982}%
  \BibitemOpen
  \bibfield  {author} {\bibinfo {author} {\bibfnamefont {N.~D.}\ \bibnamefont
  {Birrell}}\ and\ \bibinfo {author} {\bibfnamefont {P.~C.~W.}\ \bibnamefont
  {Davies}},\ }\href {\doibase 10.1017/CBO9780511622632} {\emph {\bibinfo
  {title} {{Quantum Fields in Curved Space}}}},\ Cambridge Monographs on
  Mathematical Physics\ (\bibinfo  {publisher} {Cambridge Univ. Press},\
  \bibinfo {address} {Cambridge, UK},\ \bibinfo {year} {1984})\BibitemShut
  {NoStop}%
\bibitem [{\citenamefont {Wald}(1994)}]{Wald1995}%
  \BibitemOpen
  \bibfield  {author} {\bibinfo {author} {\bibfnamefont {R.}~\bibnamefont
  {Wald}},\ }\href@noop {} {\enquote {\bibinfo {title} {{Quantum Field Theory
  in Curved Spacetime and Black Hole Thermodynamics}},}\ }\bibinfo
  {howpublished} {University of Chicago Press} (\bibinfo {year}
  {1994})\BibitemShut {NoStop}%
\bibitem [{\citenamefont {Hawking}(1976)}]{Hawking1976}%
  \BibitemOpen
  \bibfield  {author} {\bibinfo {author} {\bibfnamefont {S.~W.}\ \bibnamefont
  {Hawking}},\ }\href {\doibase 10.1103/PhysRevD.14.2460} {\bibfield  {journal}
  {\bibinfo  {journal} {Phys. Rev. D}\ }\textbf {\bibinfo {volume} {14}},\
  \bibinfo {pages} {2460} (\bibinfo {year} {1976})}\BibitemShut {NoStop}%
\bibitem [{\citenamefont {Unruh}\ and\ \citenamefont {Wald}(2017)}]{Unruh2017}%
  \BibitemOpen
  \bibfield  {author} {\bibinfo {author} {\bibfnamefont {W.~G.}\ \bibnamefont
  {Unruh}}\ and\ \bibinfo {author} {\bibfnamefont {R.~M.}\ \bibnamefont
  {Wald}},\ }\href {\doibase 10.1088/1361-6633/aa778e} {\bibfield  {journal}
  {\bibinfo  {journal} {Rept. Prog. Phys.}\ }\textbf {\bibinfo {volume} {80}},\
  \bibinfo {pages} {092002} (\bibinfo {year} {2017})},\ \Eprint
  {http://arxiv.org/abs/1703.02140} {arXiv:1703.02140 [hep-th]} \BibitemShut
  {NoStop}%
\bibitem [{\citenamefont {Almheiri}\ \emph {et~al.}(2013)\citenamefont
  {Almheiri}, \citenamefont {Marolf}, \citenamefont {Polchinski},\ and\
  \citenamefont {Sully}}]{Almheiri2012}%
  \BibitemOpen
  \bibfield  {author} {\bibinfo {author} {\bibfnamefont {A.}~\bibnamefont
  {Almheiri}}, \bibinfo {author} {\bibfnamefont {D.}~\bibnamefont {Marolf}},
  \bibinfo {author} {\bibfnamefont {J.}~\bibnamefont {Polchinski}}, \ and\
  \bibinfo {author} {\bibfnamefont {J.}~\bibnamefont {Sully}},\ }\href
  {\doibase 10.1007/JHEP02(2013)062} {\bibfield  {journal} {\bibinfo  {journal}
  {JHEP}\ }\textbf {\bibinfo {volume} {02}},\ \bibinfo {pages} {062} (\bibinfo
  {year} {2013})},\ \Eprint {http://arxiv.org/abs/1207.3123} {arXiv:1207.3123
  [hep-th]} \BibitemShut {NoStop}%
\bibitem [{\citenamefont {Visser}(2003)}]{Visser2001}%
  \BibitemOpen
  \bibfield  {author} {\bibinfo {author} {\bibfnamefont {M.}~\bibnamefont
  {Visser}},\ }\href {\doibase 10.1142/S0218271803003190} {\bibfield  {journal}
  {\bibinfo  {journal} {Int. J. Mod. Phys. D}\ }\textbf {\bibinfo {volume}
  {12}},\ \bibinfo {pages} {649} (\bibinfo {year} {2003})},\ \Eprint
  {http://arxiv.org/abs/hep-th/0106111} {arXiv:hep-th/0106111} \BibitemShut
  {NoStop}%
\bibitem [{\citenamefont {Barcelo}\ \emph
  {et~al.}(2006{\natexlab{a}})\citenamefont {Barcelo}, \citenamefont
  {Liberati}, \citenamefont {Sonego},\ and\ \citenamefont
  {Visser}}]{Barcelo2006}%
  \BibitemOpen
  \bibfield  {author} {\bibinfo {author} {\bibfnamefont {C.}~\bibnamefont
  {Barcelo}}, \bibinfo {author} {\bibfnamefont {S.}~\bibnamefont {Liberati}},
  \bibinfo {author} {\bibfnamefont {S.}~\bibnamefont {Sonego}}, \ and\ \bibinfo
  {author} {\bibfnamefont {M.}~\bibnamefont {Visser}},\ }\href {\doibase
  10.1103/PhysRevLett.97.171301} {\bibfield  {journal} {\bibinfo  {journal}
  {Phys. Rev. Lett.}\ }\textbf {\bibinfo {volume} {97}},\ \bibinfo {pages}
  {171301} (\bibinfo {year} {2006}{\natexlab{a}})},\ \Eprint
  {http://arxiv.org/abs/gr-qc/0607008} {arXiv:gr-qc/0607008} \BibitemShut
  {NoStop}%
\bibitem [{\citenamefont {Barcelo}\ \emph
  {et~al.}(2006{\natexlab{b}})\citenamefont {Barcelo}, \citenamefont
  {Liberati}, \citenamefont {Sonego},\ and\ \citenamefont
  {Visser}}]{Barcelo2006b}%
  \BibitemOpen
  \bibfield  {author} {\bibinfo {author} {\bibfnamefont {C.}~\bibnamefont
  {Barcelo}}, \bibinfo {author} {\bibfnamefont {S.}~\bibnamefont {Liberati}},
  \bibinfo {author} {\bibfnamefont {S.}~\bibnamefont {Sonego}}, \ and\ \bibinfo
  {author} {\bibfnamefont {M.}~\bibnamefont {Visser}},\ }\href {\doibase
  10.1088/0264-9381/23/17/014} {\bibfield  {journal} {\bibinfo  {journal}
  {Class. Quant. Grav.}\ }\textbf {\bibinfo {volume} {23}},\ \bibinfo {pages}
  {5341} (\bibinfo {year} {2006}{\natexlab{b}})},\ \Eprint
  {http://arxiv.org/abs/gr-qc/0604058} {arXiv:gr-qc/0604058} \BibitemShut
  {NoStop}%
\bibitem [{\citenamefont {Barcelo}\ \emph {et~al.}(2005)\citenamefont
  {Barcelo}, \citenamefont {Liberati},\ and\ \citenamefont
  {Visser}}]{Barcelo2005}%
  \BibitemOpen
  \bibfield  {author} {\bibinfo {author} {\bibfnamefont {C.}~\bibnamefont
  {Barcelo}}, \bibinfo {author} {\bibfnamefont {S.}~\bibnamefont {Liberati}}, \
  and\ \bibinfo {author} {\bibfnamefont {M.}~\bibnamefont {Visser}},\ }\href
  {\doibase 10.12942/lrr-2005-12} {\bibfield  {journal} {\bibinfo  {journal}
  {Living Rev. Rel.}\ }\textbf {\bibinfo {volume} {8}},\ \bibinfo {pages} {12}
  (\bibinfo {year} {2005})},\ \Eprint {http://arxiv.org/abs/gr-qc/0505065}
  {arXiv:gr-qc/0505065} \BibitemShut {NoStop}%
\bibitem [{\citenamefont {Barcel\'o}(2019)}]{Barcelo2018}%
  \BibitemOpen
  \bibfield  {author} {\bibinfo {author} {\bibfnamefont {C.}~\bibnamefont
  {Barcel\'o}},\ }\href {\doibase 10.1038/s41567-018-0367-6} {\bibfield
  {journal} {\bibinfo  {journal} {Nature Phys.}\ }\textbf {\bibinfo {volume}
  {15}},\ \bibinfo {pages} {210} (\bibinfo {year} {2019})}\BibitemShut
  {NoStop}%
\bibitem [{\citenamefont {Garay}\ \emph {et~al.}(2000)\citenamefont {Garay},
  \citenamefont {Anglin}, \citenamefont {Cirac},\ and\ \citenamefont
  {Zoller}}]{Garay1999}%
  \BibitemOpen
  \bibfield  {author} {\bibinfo {author} {\bibfnamefont {L.~J.}\ \bibnamefont
  {Garay}}, \bibinfo {author} {\bibfnamefont {J.~R.}\ \bibnamefont {Anglin}},
  \bibinfo {author} {\bibfnamefont {J.~I.}\ \bibnamefont {Cirac}}, \ and\
  \bibinfo {author} {\bibfnamefont {P.}~\bibnamefont {Zoller}},\ }\href
  {\doibase 10.1103/PhysRevLett.85.4643} {\bibfield  {journal} {\bibinfo
  {journal} {Phys. Rev. Lett.}\ }\textbf {\bibinfo {volume} {85}},\ \bibinfo
  {pages} {4643} (\bibinfo {year} {2000})},\ \Eprint
  {http://arxiv.org/abs/gr-qc/0002015} {arXiv:gr-qc/0002015} \BibitemShut
  {NoStop}%
\bibitem [{\citenamefont {Garay}\ \emph {et~al.}(2001)\citenamefont {Garay},
  \citenamefont {Anglin}, \citenamefont {Cirac},\ and\ \citenamefont
  {Zoller}}]{Garay2000}%
  \BibitemOpen
  \bibfield  {author} {\bibinfo {author} {\bibfnamefont {L.~J.}\ \bibnamefont
  {Garay}}, \bibinfo {author} {\bibfnamefont {J.~R.}\ \bibnamefont {Anglin}},
  \bibinfo {author} {\bibfnamefont {J.~I.}\ \bibnamefont {Cirac}}, \ and\
  \bibinfo {author} {\bibfnamefont {P.}~\bibnamefont {Zoller}},\ }\href
  {\doibase 10.1103/PhysRevA.63.023611} {\bibfield  {journal} {\bibinfo
  {journal} {Phys. Rev. A}\ }\textbf {\bibinfo {volume} {63}},\ \bibinfo
  {pages} {023611} (\bibinfo {year} {2001})},\ \Eprint
  {http://arxiv.org/abs/gr-qc/0005131} {arXiv:gr-qc/0005131} \BibitemShut
  {NoStop}%
\bibitem [{\citenamefont {Barcelo}\ \emph {et~al.}(2003)\citenamefont
  {Barcelo}, \citenamefont {Liberati},\ and\ \citenamefont
  {Visser}}]{Barcelo2001}%
  \BibitemOpen
  \bibfield  {author} {\bibinfo {author} {\bibfnamefont {C.}~\bibnamefont
  {Barcelo}}, \bibinfo {author} {\bibfnamefont {S.}~\bibnamefont {Liberati}}, \
  and\ \bibinfo {author} {\bibfnamefont {M.}~\bibnamefont {Visser}},\ }\href
  {\doibase 10.1142/S0217751X0301615X} {\bibfield  {journal} {\bibinfo
  {journal} {Int. J. Mod. Phys. A}\ }\textbf {\bibinfo {volume} {18}},\
  \bibinfo {pages} {3735} (\bibinfo {year} {2003})},\ \Eprint
  {http://arxiv.org/abs/gr-qc/0110036} {arXiv:gr-qc/0110036} \BibitemShut
  {NoStop}%
\bibitem [{\citenamefont {Steinhauer}(2016)}]{Steinhauer2015}%
  \BibitemOpen
  \bibfield  {author} {\bibinfo {author} {\bibfnamefont {J.}~\bibnamefont
  {Steinhauer}},\ }\href {\doibase 10.1038/nphys3863} {\bibfield  {journal}
  {\bibinfo  {journal} {Nature Phys.}\ }\textbf {\bibinfo {volume} {12}},\
  \bibinfo {pages} {959} (\bibinfo {year} {2016})},\ \Eprint
  {http://arxiv.org/abs/1510.00621} {arXiv:1510.00621 [gr-qc]} \BibitemShut
  {NoStop}%
\bibitem [{\citenamefont {Muñoz~de Nova}\ \emph {et~al.}(2019)\citenamefont
  {Muñoz~de Nova}, \citenamefont {Golubkov}, \citenamefont {Kolobov},\ and\
  \citenamefont {et~al.}}]{Nova:2019}%
  \BibitemOpen
  \bibfield  {author} {\bibinfo {author} {\bibfnamefont {J.}~\bibnamefont
  {Muñoz~de Nova}}, \bibinfo {author} {\bibfnamefont {K.}~\bibnamefont
  {Golubkov}}, \bibinfo {author} {\bibfnamefont {V.}~\bibnamefont {Kolobov}}, \
  and\ \bibinfo {author} {\bibnamefont {et~al.}},\ }\href {\doibase
  10.1038/s41586-019-1241-0} {\bibfield  {journal} {\bibinfo  {journal}
  {Nature}\ }\textbf {\bibinfo {volume} {569}},\ \bibinfo {pages} {688}
  (\bibinfo {year} {2019})},\ \Eprint {http://arxiv.org/abs/1809.00913}
  {arXiv:1809.00913} \BibitemShut {NoStop}%
\bibitem [{\citenamefont {Kolobov}\ \emph {et~al.}(2021)\citenamefont
  {Kolobov}, \citenamefont {Golubkov}, \citenamefont {Mu\~noz~de Nova},\ and\
  \citenamefont {Steinhauer}}]{Kolobov2019}%
  \BibitemOpen
  \bibfield  {author} {\bibinfo {author} {\bibfnamefont {V.~I.}\ \bibnamefont
  {Kolobov}}, \bibinfo {author} {\bibfnamefont {K.}~\bibnamefont {Golubkov}},
  \bibinfo {author} {\bibfnamefont {J.~R.}\ \bibnamefont {Mu\~noz~de Nova}}, \
  and\ \bibinfo {author} {\bibfnamefont {J.}~\bibnamefont {Steinhauer}},\
  }\href {\doibase 10.1038/s41567-020-01076-0} {\bibfield  {journal} {\bibinfo
  {journal} {Nature Phys.}\ }\textbf {\bibinfo {volume} {17}},\ \bibinfo
  {pages} {362} (\bibinfo {year} {2021})},\ \Eprint
  {http://arxiv.org/abs/1910.09363} {arXiv:1910.09363 [gr-qc]} \BibitemShut
  {NoStop}%
\bibitem [{\citenamefont {Isoard}(2020)}]{IsoardThesis}%
  \BibitemOpen
  \bibfield  {author} {\bibinfo {author} {\bibfnamefont {M.}~\bibnamefont
  {Isoard}},\ }\href@noop {} {\enquote {\bibinfo {title} {{Theoretical study of
  quantum correlations and nonlinear fluctuations in quantum gases}},}\
  }\bibinfo {howpublished} {Phd thesis, University of Paris-Saclay} (\bibinfo
  {year} {2020})\BibitemShut {NoStop}%
\bibitem [{\citenamefont {Leonhardt}(2018)}]{Leonhardt2016}%
  \BibitemOpen
  \bibfield  {author} {\bibinfo {author} {\bibfnamefont {U.}~\bibnamefont
  {Leonhardt}},\ }\href {\doibase 10.1002/andp.201700114} {\bibfield  {journal}
  {\bibinfo  {journal} {Annalen Phys.}\ }\textbf {\bibinfo {volume} {530}},\
  \bibinfo {pages} {1700114} (\bibinfo {year} {2018})},\ \Eprint
  {http://arxiv.org/abs/1609.03803} {arXiv:1609.03803 [gr-qc]} \BibitemShut
  {NoStop}%
\bibitem [{\citenamefont {Steinhauer}(2018)}]{Steinhauer2018}%
  \BibitemOpen
  \bibfield  {author} {\bibinfo {author} {\bibfnamefont {J.}~\bibnamefont
  {Steinhauer}},\ }\href {\doibase 10.1002/andp.201700459} {\bibfield
  {journal} {\bibinfo  {journal} {Annalen Phys.}\ }\textbf {\bibinfo {volume}
  {530}},\ \bibinfo {pages} {1700459} (\bibinfo {year} {2018})}\BibitemShut
  {NoStop}%
\bibitem [{\citenamefont {Schutzhold}\ and\ \citenamefont
  {Unruh}(2002)}]{Schutzhold2002}%
  \BibitemOpen
  \bibfield  {author} {\bibinfo {author} {\bibfnamefont {R.}~\bibnamefont
  {Schutzhold}}\ and\ \bibinfo {author} {\bibfnamefont {W.~G.}\ \bibnamefont
  {Unruh}},\ }\href {\doibase 10.1103/PhysRevD.66.044019} {\bibfield  {journal}
  {\bibinfo  {journal} {Phys. Rev. D}\ }\textbf {\bibinfo {volume} {66}},\
  \bibinfo {pages} {044019} (\bibinfo {year} {2002})},\ \Eprint
  {http://arxiv.org/abs/gr-qc/0205099} {arXiv:gr-qc/0205099} \BibitemShut
  {NoStop}%
\bibitem [{\citenamefont {Rousseaux}\ \emph {et~al.}(2008)\citenamefont
  {Rousseaux}, \citenamefont {Mathis}, \citenamefont {Maissa}, \citenamefont
  {Philbin},\ and\ \citenamefont {Leonhardt}}]{Rousseaux2007}%
  \BibitemOpen
  \bibfield  {author} {\bibinfo {author} {\bibfnamefont {G.}~\bibnamefont
  {Rousseaux}}, \bibinfo {author} {\bibfnamefont {C.}~\bibnamefont {Mathis}},
  \bibinfo {author} {\bibfnamefont {P.}~\bibnamefont {Maissa}}, \bibinfo
  {author} {\bibfnamefont {T.~G.}\ \bibnamefont {Philbin}}, \ and\ \bibinfo
  {author} {\bibfnamefont {U.}~\bibnamefont {Leonhardt}},\ }\href {\doibase
  10.1088/1367-2630/10/5/053015} {\bibfield  {journal} {\bibinfo  {journal}
  {New J. Phys.}\ }\textbf {\bibinfo {volume} {10}},\ \bibinfo {pages} {053015}
  (\bibinfo {year} {2008})},\ \Eprint {http://arxiv.org/abs/0711.4767}
  {arXiv:0711.4767 [gr-qc]} \BibitemShut {NoStop}%
\bibitem [{\citenamefont {Weinfurtner}\ \emph {et~al.}(2011)\citenamefont
  {Weinfurtner}, \citenamefont {Tedford}, \citenamefont {Penrice},
  \citenamefont {Unruh},\ and\ \citenamefont {Lawrence}}]{Weinfurtner2010}%
  \BibitemOpen
  \bibfield  {author} {\bibinfo {author} {\bibfnamefont {S.}~\bibnamefont
  {Weinfurtner}}, \bibinfo {author} {\bibfnamefont {E.~W.}\ \bibnamefont
  {Tedford}}, \bibinfo {author} {\bibfnamefont {M.~C.~J.}\ \bibnamefont
  {Penrice}}, \bibinfo {author} {\bibfnamefont {W.~G.}\ \bibnamefont {Unruh}},
  \ and\ \bibinfo {author} {\bibfnamefont {G.~A.}\ \bibnamefont {Lawrence}},\
  }\href {\doibase 10.1103/PhysRevLett.106.021302} {\bibfield  {journal}
  {\bibinfo  {journal} {Phys. Rev. Lett.}\ }\textbf {\bibinfo {volume} {106}},\
  \bibinfo {pages} {021302} (\bibinfo {year} {2011})},\ \Eprint
  {http://arxiv.org/abs/1008.1911} {arXiv:1008.1911 [gr-qc]} \BibitemShut
  {NoStop}%
\bibitem [{\citenamefont {Weinfurtner}\ \emph {et~al.}(2013)\citenamefont
  {Weinfurtner}, \citenamefont {Tedford}, \citenamefont {Penrice},
  \citenamefont {Unruh},\ and\ \citenamefont {Lawrence}}]{Weinfurtner2013}%
  \BibitemOpen
  \bibfield  {author} {\bibinfo {author} {\bibfnamefont {S.}~\bibnamefont
  {Weinfurtner}}, \bibinfo {author} {\bibfnamefont {E.~W.}\ \bibnamefont
  {Tedford}}, \bibinfo {author} {\bibfnamefont {M.~C.~J.}\ \bibnamefont
  {Penrice}}, \bibinfo {author} {\bibfnamefont {W.~G.}\ \bibnamefont {Unruh}},
  \ and\ \bibinfo {author} {\bibfnamefont {G.~A.}\ \bibnamefont {Lawrence}},\
  }\href {\doibase 10.1007/978-3-319-00266-8_8} {\bibfield  {journal} {\bibinfo
   {journal} {Lect. Notes Phys.}\ }\textbf {\bibinfo {volume} {870}},\ \bibinfo
  {pages} {167} (\bibinfo {year} {2013})},\ \Eprint
  {http://arxiv.org/abs/1302.0375} {arXiv:1302.0375 [gr-qc]} \BibitemShut
  {NoStop}%
\bibitem [{\citenamefont {Michel}\ and\ \citenamefont
  {Parentani}(2014)}]{Michel2014}%
  \BibitemOpen
  \bibfield  {author} {\bibinfo {author} {\bibfnamefont {F.}~\bibnamefont
  {Michel}}\ and\ \bibinfo {author} {\bibfnamefont {R.}~\bibnamefont
  {Parentani}},\ }\href {\doibase 10.1103/PhysRevD.90.044033} {\bibfield
  {journal} {\bibinfo  {journal} {Phys. Rev. D}\ }\textbf {\bibinfo {volume}
  {90}},\ \bibinfo {pages} {044033} (\bibinfo {year} {2014})},\ \Eprint
  {http://arxiv.org/abs/1404.7482} {arXiv:1404.7482 [gr-qc]} \BibitemShut
  {NoStop}%
\bibitem [{\citenamefont {Coutant}\ and\ \citenamefont
  {Weinfurtner}(2016)}]{Coutant2016}%
  \BibitemOpen
  \bibfield  {author} {\bibinfo {author} {\bibfnamefont {A.}~\bibnamefont
  {Coutant}}\ and\ \bibinfo {author} {\bibfnamefont {S.}~\bibnamefont
  {Weinfurtner}},\ }\href {\doibase 10.1103/PhysRevD.94.064026} {\bibfield
  {journal} {\bibinfo  {journal} {Phys. Rev. D}\ }\textbf {\bibinfo {volume}
  {94}},\ \bibinfo {pages} {064026} (\bibinfo {year} {2016})},\ \Eprint
  {http://arxiv.org/abs/1603.02746} {arXiv:1603.02746 [gr-qc]} \BibitemShut
  {NoStop}%
\bibitem [{\citenamefont {Euv\'e}\ \emph {et~al.}(2015)\citenamefont {Euv\'e},
  \citenamefont {Michel}, \citenamefont {Parentani},\ and\ \citenamefont
  {Rousseaux}}]{Euve2014}%
  \BibitemOpen
  \bibfield  {author} {\bibinfo {author} {\bibfnamefont {L.-P.}\ \bibnamefont
  {Euv\'e}}, \bibinfo {author} {\bibfnamefont {F.}~\bibnamefont {Michel}},
  \bibinfo {author} {\bibfnamefont {R.}~\bibnamefont {Parentani}}, \ and\
  \bibinfo {author} {\bibfnamefont {G.}~\bibnamefont {Rousseaux}},\ }\href
  {\doibase 10.1103/PhysRevD.91.024020} {\bibfield  {journal} {\bibinfo
  {journal} {Phys. Rev. D}\ }\textbf {\bibinfo {volume} {91}},\ \bibinfo
  {pages} {024020} (\bibinfo {year} {2015})},\ \Eprint
  {http://arxiv.org/abs/1409.3830} {arXiv:1409.3830 [gr-qc]} \BibitemShut
  {NoStop}%
\bibitem [{\citenamefont {Euv\'e}\ \emph {et~al.}(2020)\citenamefont {Euv\'e},
  \citenamefont {Robertson}, \citenamefont {James}, \citenamefont {Fabbri},\
  and\ \citenamefont {Rousseaux}}]{Euve2018}%
  \BibitemOpen
  \bibfield  {author} {\bibinfo {author} {\bibfnamefont {L.-P.}\ \bibnamefont
  {Euv\'e}}, \bibinfo {author} {\bibfnamefont {S.}~\bibnamefont {Robertson}},
  \bibinfo {author} {\bibfnamefont {N.}~\bibnamefont {James}}, \bibinfo
  {author} {\bibfnamefont {A.}~\bibnamefont {Fabbri}}, \ and\ \bibinfo {author}
  {\bibfnamefont {G.}~\bibnamefont {Rousseaux}},\ }\href {\doibase
  10.1103/PhysRevLett.124.141101} {\bibfield  {journal} {\bibinfo  {journal}
  {Phys. Rev. Lett.}\ }\textbf {\bibinfo {volume} {124}},\ \bibinfo {pages}
  {141101} (\bibinfo {year} {2020})},\ \Eprint
  {http://arxiv.org/abs/1806.05539} {arXiv:1806.05539 [gr-qc]} \BibitemShut
  {NoStop}%
\bibitem [{\citenamefont {Schutzhold}\ and\ \citenamefont
  {Unruh}(2005)}]{Schutzhold:2004tv}%
  \BibitemOpen
  \bibfield  {author} {\bibinfo {author} {\bibfnamefont {R.}~\bibnamefont
  {Schutzhold}}\ and\ \bibinfo {author} {\bibfnamefont {W.~G.}\ \bibnamefont
  {Unruh}},\ }\href {\doibase 10.1103/PhysRevLett.95.031301} {\bibfield
  {journal} {\bibinfo  {journal} {Phys. Rev. Lett.}\ }\textbf {\bibinfo
  {volume} {95}},\ \bibinfo {pages} {031301} (\bibinfo {year} {2005})},\
  \Eprint {http://arxiv.org/abs/quant-ph/0408145} {arXiv:quant-ph/0408145}
  \BibitemShut {NoStop}%
\bibitem [{\citenamefont {Philbin}\ \emph {et~al.}(2008)\citenamefont
  {Philbin}, \citenamefont {Kuklewicz}, \citenamefont {Robertson},
  \citenamefont {Hill}, \citenamefont {Konig},\ and\ \citenamefont
  {Leonhardt}}]{Philbin:2007ji}%
  \BibitemOpen
  \bibfield  {author} {\bibinfo {author} {\bibfnamefont {T.~G.}\ \bibnamefont
  {Philbin}}, \bibinfo {author} {\bibfnamefont {C.}~\bibnamefont {Kuklewicz}},
  \bibinfo {author} {\bibfnamefont {S.}~\bibnamefont {Robertson}}, \bibinfo
  {author} {\bibfnamefont {S.}~\bibnamefont {Hill}}, \bibinfo {author}
  {\bibfnamefont {F.}~\bibnamefont {Konig}}, \ and\ \bibinfo {author}
  {\bibfnamefont {U.}~\bibnamefont {Leonhardt}},\ }\href {\doibase
  10.1126/science.1153625} {\bibfield  {journal} {\bibinfo  {journal}
  {Science}\ }\textbf {\bibinfo {volume} {319}},\ \bibinfo {pages} {1367}
  (\bibinfo {year} {2008})},\ \Eprint {http://arxiv.org/abs/0711.4796}
  {arXiv:0711.4796 [gr-qc]} \BibitemShut {NoStop}%
\bibitem [{\citenamefont {Belgiorno}\ \emph {et~al.}(2010)\citenamefont
  {Belgiorno}, \citenamefont {Cacciatori}, \citenamefont {Ortenzi},
  \citenamefont {Sala},\ and\ \citenamefont {Faccio}}]{Belgiorno2010}%
  \BibitemOpen
  \bibfield  {author} {\bibinfo {author} {\bibfnamefont {F.}~\bibnamefont
  {Belgiorno}}, \bibinfo {author} {\bibfnamefont {S.~L.}\ \bibnamefont
  {Cacciatori}}, \bibinfo {author} {\bibfnamefont {G.}~\bibnamefont {Ortenzi}},
  \bibinfo {author} {\bibfnamefont {V.~G.}\ \bibnamefont {Sala}}, \ and\
  \bibinfo {author} {\bibfnamefont {D.}~\bibnamefont {Faccio}},\ }\href
  {\doibase 10.1103/PhysRevLett.104.140403} {\bibfield  {journal} {\bibinfo
  {journal} {Phys. Rev. Lett.}\ }\textbf {\bibinfo {volume} {104}},\ \bibinfo
  {pages} {140403} (\bibinfo {year} {2010})}\BibitemShut {NoStop}%
\bibitem [{\citenamefont {Rubino}\ \emph {et~al.}(2011)\citenamefont {Rubino},
  \citenamefont {Belgiorno}, \citenamefont {Cacciatori}, \citenamefont
  {Clerici}, \citenamefont {Gorini}, \citenamefont {Ortenzi}, \citenamefont
  {Rizzi}, \citenamefont {Sala}, \citenamefont {Kolesik},\ and\ \citenamefont
  {Faccio}}]{Rubino2011}%
  \BibitemOpen
  \bibfield  {author} {\bibinfo {author} {\bibfnamefont {E.}~\bibnamefont
  {Rubino}}, \bibinfo {author} {\bibfnamefont {F.}~\bibnamefont {Belgiorno}},
  \bibinfo {author} {\bibfnamefont {S.~L.}\ \bibnamefont {Cacciatori}},
  \bibinfo {author} {\bibfnamefont {M.}~\bibnamefont {Clerici}}, \bibinfo
  {author} {\bibfnamefont {V.}~\bibnamefont {Gorini}}, \bibinfo {author}
  {\bibfnamefont {G.}~\bibnamefont {Ortenzi}}, \bibinfo {author} {\bibfnamefont
  {L.}~\bibnamefont {Rizzi}}, \bibinfo {author} {\bibfnamefont {V.~G.}\
  \bibnamefont {Sala}}, \bibinfo {author} {\bibfnamefont {M.}~\bibnamefont
  {Kolesik}}, \ and\ \bibinfo {author} {\bibfnamefont {D.}~\bibnamefont
  {Faccio}},\ }\href {\doibase 10.1088/1367-2630/13/8/085005} {\bibfield
  {journal} {\bibinfo  {journal} {New J. Phys.}\ }\textbf {\bibinfo {volume}
  {13}},\ \bibinfo {pages} {085005} (\bibinfo {year} {2011})}\BibitemShut
  {NoStop}%
\bibitem [{\citenamefont {Drori}\ \emph {et~al.}(2019)\citenamefont {Drori},
  \citenamefont {Rosenberg}, \citenamefont {Bermudez}, \citenamefont
  {Silberberg},\ and\ \citenamefont {Leonhardt}}]{Drori2018}%
  \BibitemOpen
  \bibfield  {author} {\bibinfo {author} {\bibfnamefont {J.}~\bibnamefont
  {Drori}}, \bibinfo {author} {\bibfnamefont {Y.}~\bibnamefont {Rosenberg}},
  \bibinfo {author} {\bibfnamefont {D.}~\bibnamefont {Bermudez}}, \bibinfo
  {author} {\bibfnamefont {Y.}~\bibnamefont {Silberberg}}, \ and\ \bibinfo
  {author} {\bibfnamefont {U.}~\bibnamefont {Leonhardt}},\ }\href {\doibase
  10.1103/PhysRevLett.122.010404} {\bibfield  {journal} {\bibinfo  {journal}
  {Phys. Rev. Lett.}\ }\textbf {\bibinfo {volume} {122}},\ \bibinfo {pages}
  {010404} (\bibinfo {year} {2019})},\ \Eprint
  {http://arxiv.org/abs/1808.09244} {arXiv:1808.09244 [gr-qc]} \BibitemShut
  {NoStop}%
\bibitem [{\citenamefont {Carusotto}\ and\ \citenamefont
  {Ciuti}(2004)}]{Carusotto2004}%
  \BibitemOpen
  \bibfield  {author} {\bibinfo {author} {\bibfnamefont {I.}~\bibnamefont
  {Carusotto}}\ and\ \bibinfo {author} {\bibfnamefont {C.}~\bibnamefont
  {Ciuti}},\ }\href {\doibase 10.1103/PhysRevLett.93.166401} {\bibfield
  {journal} {\bibinfo  {journal} {Phys. Rev. Lett.}\ }\textbf {\bibinfo
  {volume} {93}},\ \bibinfo {pages} {166401} (\bibinfo {year}
  {2004})}\BibitemShut {NoStop}%
\bibitem [{\citenamefont {Nguyen}\ \emph {et~al.}(2015)\citenamefont {Nguyen},
  \citenamefont {Gerace}, \citenamefont {Carusotto}, \citenamefont {Sanvitto},
  \citenamefont {Galopin}, \citenamefont {Lema\^{\i}tre}, \citenamefont
  {Sagnes}, \citenamefont {Bloch},\ and\ \citenamefont {Amo}}]{Nguyen2015}%
  \BibitemOpen
  \bibfield  {author} {\bibinfo {author} {\bibfnamefont {H.~S.}\ \bibnamefont
  {Nguyen}}, \bibinfo {author} {\bibfnamefont {D.}~\bibnamefont {Gerace}},
  \bibinfo {author} {\bibfnamefont {I.}~\bibnamefont {Carusotto}}, \bibinfo
  {author} {\bibfnamefont {D.}~\bibnamefont {Sanvitto}}, \bibinfo {author}
  {\bibfnamefont {E.}~\bibnamefont {Galopin}}, \bibinfo {author} {\bibfnamefont
  {A.}~\bibnamefont {Lema\^{\i}tre}}, \bibinfo {author} {\bibfnamefont
  {I.}~\bibnamefont {Sagnes}}, \bibinfo {author} {\bibfnamefont
  {J.}~\bibnamefont {Bloch}}, \ and\ \bibinfo {author} {\bibfnamefont
  {A.}~\bibnamefont {Amo}},\ }\href {\doibase 10.1103/PhysRevLett.114.036402}
  {\bibfield  {journal} {\bibinfo  {journal} {Phys. Rev. Lett.}\ }\textbf
  {\bibinfo {volume} {114}},\ \bibinfo {pages} {036402} (\bibinfo {year}
  {2015})}\BibitemShut {NoStop}%
\bibitem [{\citenamefont {Stephens}\ \emph {et~al.}(1994)\citenamefont
  {Stephens}, \citenamefont {'t~Hooft},\ and\ \citenamefont
  {Whiting}}]{Stephens1993}%
  \BibitemOpen
  \bibfield  {author} {\bibinfo {author} {\bibfnamefont {C.~R.}\ \bibnamefont
  {Stephens}}, \bibinfo {author} {\bibfnamefont {G.}~\bibnamefont {'t~Hooft}},
  \ and\ \bibinfo {author} {\bibfnamefont {B.~F.}\ \bibnamefont {Whiting}},\
  }\href {\doibase 10.1088/0264-9381/11/3/014} {\bibfield  {journal} {\bibinfo
  {journal} {Class. Quant. Grav.}\ }\textbf {\bibinfo {volume} {11}},\ \bibinfo
  {pages} {621} (\bibinfo {year} {1994})},\ \Eprint
  {http://arxiv.org/abs/gr-qc/9310006} {arXiv:gr-qc/9310006} \BibitemShut
  {NoStop}%
\bibitem [{\citenamefont {Barcelo}\ \emph {et~al.}(2015)\citenamefont
  {Barcelo}, \citenamefont {Carballo-Rubio}, \citenamefont {Garay},\ and\
  \citenamefont {Jannes}}]{Barcelo2014}%
  \BibitemOpen
  \bibfield  {author} {\bibinfo {author} {\bibfnamefont {C.}~\bibnamefont
  {Barcelo}}, \bibinfo {author} {\bibfnamefont {R.}~\bibnamefont
  {Carballo-Rubio}}, \bibinfo {author} {\bibfnamefont {L.~J.}\ \bibnamefont
  {Garay}}, \ and\ \bibinfo {author} {\bibfnamefont {G.}~\bibnamefont
  {Jannes}},\ }\href {\doibase 10.1088/0264-9381/32/3/035012} {\bibfield
  {journal} {\bibinfo  {journal} {Class. Quant. Grav.}\ }\textbf {\bibinfo
  {volume} {32}},\ \bibinfo {pages} {035012} (\bibinfo {year} {2015})},\
  \Eprint {http://arxiv.org/abs/1409.1501} {arXiv:1409.1501 [gr-qc]}
  \BibitemShut {NoStop}%
\bibitem [{\citenamefont {Husain}\ \emph
  {et~al.}(2022{\natexlab{a}})\citenamefont {Husain}, \citenamefont {Kelly},
  \citenamefont {Santacruz},\ and\ \citenamefont {Wilson-Ewing}}]{Husain2021}%
  \BibitemOpen
  \bibfield  {author} {\bibinfo {author} {\bibfnamefont {V.}~\bibnamefont
  {Husain}}, \bibinfo {author} {\bibfnamefont {J.~G.}\ \bibnamefont {Kelly}},
  \bibinfo {author} {\bibfnamefont {R.}~\bibnamefont {Santacruz}}, \ and\
  \bibinfo {author} {\bibfnamefont {E.}~\bibnamefont {Wilson-Ewing}},\ }\href
  {\doibase 10.1103/PhysRevLett.128.121301} {\bibfield  {journal} {\bibinfo
  {journal} {Phys. Rev. Lett.}\ }\textbf {\bibinfo {volume} {128}},\ \bibinfo
  {pages} {121301} (\bibinfo {year} {2022}{\natexlab{a}})},\ \Eprint
  {http://arxiv.org/abs/2109.08667} {arXiv:2109.08667 [gr-qc]} \BibitemShut
  {NoStop}%
\bibitem [{\citenamefont {Husain}\ \emph
  {et~al.}(2022{\natexlab{b}})\citenamefont {Husain}, \citenamefont {Kelly},
  \citenamefont {Santacruz},\ and\ \citenamefont {Wilson-Ewing}}]{Husain2022}%
  \BibitemOpen
  \bibfield  {author} {\bibinfo {author} {\bibfnamefont {V.}~\bibnamefont
  {Husain}}, \bibinfo {author} {\bibfnamefont {J.~G.}\ \bibnamefont {Kelly}},
  \bibinfo {author} {\bibfnamefont {R.}~\bibnamefont {Santacruz}}, \ and\
  \bibinfo {author} {\bibfnamefont {E.}~\bibnamefont {Wilson-Ewing}},\ }\href
  {\doibase 10.1103/PhysRevD.106.024014} {\bibfield  {journal} {\bibinfo
  {journal} {Phys. Rev. D}\ }\textbf {\bibinfo {volume} {106}},\ \bibinfo
  {pages} {024014} (\bibinfo {year} {2022}{\natexlab{b}})},\ \Eprint
  {http://arxiv.org/abs/2203.04238} {arXiv:2203.04238 [gr-qc]} \BibitemShut
  {NoStop}%
\bibitem [{\citenamefont {Kelly}\ \emph {et~al.}(2021)\citenamefont {Kelly},
  \citenamefont {Santacruz},\ and\ \citenamefont {Wilson-Ewing}}]{Kelly2020}%
  \BibitemOpen
  \bibfield  {author} {\bibinfo {author} {\bibfnamefont {J.~G.}\ \bibnamefont
  {Kelly}}, \bibinfo {author} {\bibfnamefont {R.}~\bibnamefont {Santacruz}}, \
  and\ \bibinfo {author} {\bibfnamefont {E.}~\bibnamefont {Wilson-Ewing}},\
  }\href {\doibase 10.1088/1361-6382/abd3e2} {\bibfield  {journal} {\bibinfo
  {journal} {Class. Quant. Grav.}\ }\textbf {\bibinfo {volume} {38}},\ \bibinfo
  {pages} {04LT01} (\bibinfo {year} {2021})},\ \Eprint
  {http://arxiv.org/abs/2006.09325} {arXiv:2006.09325 [gr-qc]} \BibitemShut
  {NoStop}%
\bibitem [{\citenamefont {Agullo}\ \emph {et~al.}(2022)\citenamefont {Agullo},
  \citenamefont {Brady},\ and\ \citenamefont {Kranas}}]{Agullo2021}%
  \BibitemOpen
  \bibfield  {author} {\bibinfo {author} {\bibfnamefont {I.}~\bibnamefont
  {Agullo}}, \bibinfo {author} {\bibfnamefont {A.~J.}\ \bibnamefont {Brady}}, \
  and\ \bibinfo {author} {\bibfnamefont {D.}~\bibnamefont {Kranas}},\ }\href
  {\doibase 10.1103/PhysRevLett.128.091301} {\bibfield  {journal} {\bibinfo
  {journal} {Phys. Rev. Lett.}\ }\textbf {\bibinfo {volume} {128}},\ \bibinfo
  {pages} {091301} (\bibinfo {year} {2022})},\ \Eprint
  {http://arxiv.org/abs/2107.10217} {arXiv:2107.10217 [gr-qc]} \BibitemShut
  {NoStop}%
\bibitem [{\citenamefont {Brady}\ \emph {et~al.}(2022)\citenamefont {Brady},
  \citenamefont {Agullo},\ and\ \citenamefont {Kranas}}]{Brady:2022ffk}%
  \BibitemOpen
  \bibfield  {author} {\bibinfo {author} {\bibfnamefont {A.~J.}\ \bibnamefont
  {Brady}}, \bibinfo {author} {\bibfnamefont {I.}~\bibnamefont {Agullo}}, \
  and\ \bibinfo {author} {\bibfnamefont {D.}~\bibnamefont {Kranas}},\
  }\href@noop {} {\  (\bibinfo {year} {2022})},\ \Eprint
  {http://arxiv.org/abs/2209.11317} {arXiv:2209.11317 [gr-qc]} \BibitemShut
  {NoStop}%
\bibitem [{\citenamefont {Sab\'in}(2016)}]{Sab_n_2016}%
  \BibitemOpen
  \bibfield  {author} {\bibinfo {author} {\bibfnamefont {C.}~\bibnamefont
  {Sab\'in}},\ }\href {\doibase 10.1103/physrevd.94.081501} {\bibfield
  {journal} {\bibinfo  {journal} {Physical Review D}\ }\textbf {\bibinfo
  {volume} {94}} (\bibinfo {year} {2016}),\
  10.1103/physrevd.94.081501}\BibitemShut {NoStop}%
\bibitem [{\citenamefont {Terrones}\ and\ \citenamefont
  {Sab\'in}(2021)}]{Terrones_2021}%
  \BibitemOpen
  \bibfield  {author} {\bibinfo {author} {\bibfnamefont {A.}~\bibnamefont
  {Terrones}}\ and\ \bibinfo {author} {\bibfnamefont {C.}~\bibnamefont
  {Sab\'in}},\ }\href {\doibase 10.3390/universe7120499} {\bibfield  {journal}
  {\bibinfo  {journal} {Universe}\ }\textbf {\bibinfo {volume} {7}},\ \bibinfo
  {pages} {499} (\bibinfo {year} {2021})}\BibitemShut {NoStop}%
\bibitem [{\citenamefont {Johansson}\ \emph {et~al.}(2009)\citenamefont
  {Johansson}, \citenamefont {Johansson}, \citenamefont {Wilson},\ and\
  \citenamefont {Nori}}]{Johansson:2009zz}%
  \BibitemOpen
  \bibfield  {author} {\bibinfo {author} {\bibfnamefont {J.~R.}\ \bibnamefont
  {Johansson}}, \bibinfo {author} {\bibfnamefont {G.}~\bibnamefont
  {Johansson}}, \bibinfo {author} {\bibfnamefont {C.~M.}\ \bibnamefont
  {Wilson}}, \ and\ \bibinfo {author} {\bibfnamefont {F.}~\bibnamefont
  {Nori}},\ }\href {\doibase 10.1103/PhysRevLett.103.147003} {\bibfield
  {journal} {\bibinfo  {journal} {Phys. Rev. Lett.}\ }\textbf {\bibinfo
  {volume} {103}},\ \bibinfo {pages} {147003} (\bibinfo {year} {2009})},\
  \Eprint {http://arxiv.org/abs/0906.3127} {arXiv:0906.3127
  [cond-mat.supr-con]} \BibitemShut {NoStop}%
\bibitem [{\citenamefont {Good}\ \emph {et~al.}(2013)\citenamefont {Good},
  \citenamefont {Anderson},\ and\ \citenamefont {Evans}}]{Good:2013lca}%
  \BibitemOpen
  \bibfield  {author} {\bibinfo {author} {\bibfnamefont {M.~R.~R.}\
  \bibnamefont {Good}}, \bibinfo {author} {\bibfnamefont {P.~R.}\ \bibnamefont
  {Anderson}}, \ and\ \bibinfo {author} {\bibfnamefont {C.~R.}\ \bibnamefont
  {Evans}},\ }\href {\doibase 10.1103/PhysRevD.88.025023} {\bibfield  {journal}
  {\bibinfo  {journal} {Phys. Rev. D}\ }\textbf {\bibinfo {volume} {88}},\
  \bibinfo {pages} {025023} (\bibinfo {year} {2013})},\ \Eprint
  {http://arxiv.org/abs/1303.6756} {arXiv:1303.6756 [gr-qc]} \BibitemShut
  {NoStop}%
\bibitem [{\citenamefont {Good}\ \emph {et~al.}(2020)\citenamefont {Good},
  \citenamefont {Linder},\ and\ \citenamefont {Wilczek}}]{Good:2019tnf}%
  \BibitemOpen
  \bibfield  {author} {\bibinfo {author} {\bibfnamefont {M.~R.~R.}\
  \bibnamefont {Good}}, \bibinfo {author} {\bibfnamefont {E.~V.}\ \bibnamefont
  {Linder}}, \ and\ \bibinfo {author} {\bibfnamefont {F.}~\bibnamefont
  {Wilczek}},\ }\href {\doibase 10.1103/PhysRevD.101.025012} {\bibfield
  {journal} {\bibinfo  {journal} {Phys. Rev. D}\ }\textbf {\bibinfo {volume}
  {101}},\ \bibinfo {pages} {025012} (\bibinfo {year} {2020})},\ \Eprint
  {http://arxiv.org/abs/1909.01129} {arXiv:1909.01129 [gr-qc]} \BibitemShut
  {NoStop}%
\bibitem [{\citenamefont {Johansson}\ \emph {et~al.}(2010)\citenamefont
  {Johansson}, \citenamefont {Johansson}, \citenamefont {Wilson},\ and\
  \citenamefont {Nori}}]{Johansson2010}%
  \BibitemOpen
  \bibfield  {author} {\bibinfo {author} {\bibfnamefont {J.~R.}\ \bibnamefont
  {Johansson}}, \bibinfo {author} {\bibfnamefont {G.}~\bibnamefont
  {Johansson}}, \bibinfo {author} {\bibfnamefont {C.~M.}\ \bibnamefont
  {Wilson}}, \ and\ \bibinfo {author} {\bibfnamefont {F.}~\bibnamefont
  {Nori}},\ }\href {\doibase 10.1103/PhysRevA.82.052509} {\bibfield  {journal}
  {\bibinfo  {journal} {Phys. Rev. A}\ }\textbf {\bibinfo {volume} {82}},\
  \bibinfo {pages} {052509} (\bibinfo {year} {2010})},\ \Eprint
  {http://arxiv.org/abs/1007.1058} {arXiv:1007.1058 [quant-ph]} \BibitemShut
  {NoStop}%
\bibitem [{\citenamefont {Bosco}\ \emph {et~al.}(2019)\citenamefont {Bosco},
  \citenamefont {Lindkvist},\ and\ \citenamefont {Johansson}}]{Bosco:2019ayk}%
  \BibitemOpen
  \bibfield  {author} {\bibinfo {author} {\bibfnamefont {S.}~\bibnamefont
  {Bosco}}, \bibinfo {author} {\bibfnamefont {J.}~\bibnamefont {Lindkvist}}, \
  and\ \bibinfo {author} {\bibfnamefont {G.}~\bibnamefont {Johansson}},\ }\href
  {\doibase 10.1103/PhysRevA.100.023817} {\bibfield  {journal} {\bibinfo
  {journal} {Phys. Rev. A}\ }\textbf {\bibinfo {volume} {100}},\ \bibinfo
  {pages} {023817} (\bibinfo {year} {2019})},\ \Eprint
  {http://arxiv.org/abs/1906.07966} {arXiv:1906.07966 [quant-ph]} \BibitemShut
  {NoStop}%
\bibitem [{\citenamefont {Wilson}\ \emph {et~al.}(2011)\citenamefont {Wilson},
  \citenamefont {Johansson}, \citenamefont {Pourkabirian}, \citenamefont
  {Simoen}, \citenamefont {Johansson}, \citenamefont {Duty}, \citenamefont
  {Nori},\ and\ \citenamefont {Delsing}}]{Johansson2011nat}%
  \BibitemOpen
  \bibfield  {author} {\bibinfo {author} {\bibfnamefont {C.~M.}\ \bibnamefont
  {Wilson}}, \bibinfo {author} {\bibfnamefont {G.}~\bibnamefont {Johansson}},
  \bibinfo {author} {\bibfnamefont {A.}~\bibnamefont {Pourkabirian}}, \bibinfo
  {author} {\bibfnamefont {M.}~\bibnamefont {Simoen}}, \bibinfo {author}
  {\bibfnamefont {J.~R.}\ \bibnamefont {Johansson}}, \bibinfo {author}
  {\bibfnamefont {T.}~\bibnamefont {Duty}}, \bibinfo {author} {\bibfnamefont
  {F.}~\bibnamefont {Nori}}, \ and\ \bibinfo {author} {\bibfnamefont
  {P.}~\bibnamefont {Delsing}},\ }\href {\doibase 10.1038/nature10561}
  {\bibfield  {journal} {\bibinfo  {journal} {Nature}\ }\textbf {\bibinfo
  {volume} {479}},\ \bibinfo {pages} {376–379} (\bibinfo {year}
  {2011})}\BibitemShut {NoStop}%
\bibitem [{\citenamefont {Candelas}\ and\ \citenamefont
  {Raine}(1976)}]{Candelas1976}%
  \BibitemOpen
  \bibfield  {author} {\bibinfo {author} {\bibfnamefont {P.}~\bibnamefont
  {Candelas}}\ and\ \bibinfo {author} {\bibfnamefont {D.~J.}\ \bibnamefont
  {Raine}},\ }\href {\doibase 10.1063/1.522850} {\bibfield  {journal} {\bibinfo
   {journal} {J. Math. Phys.}\ }\textbf {\bibinfo {volume} {17}},\ \bibinfo
  {pages} {2101} (\bibinfo {year} {1976})}\BibitemShut {NoStop}%
\bibitem [{\citenamefont {Davies}\ and\ \citenamefont
  {Fulling}(1976)}]{Davies1976}%
  \BibitemOpen
  \bibfield  {author} {\bibinfo {author} {\bibfnamefont {P.~C.~W.}\
  \bibnamefont {Davies}}\ and\ \bibinfo {author} {\bibfnamefont {S.~A.}\
  \bibnamefont {Fulling}},\ }\href@noop {} {\bibfield  {journal} {\bibinfo
  {journal} {Proc. Roy. Soc. Lond. A}\ }\textbf {\bibinfo {volume} {348}},\
  \bibinfo {pages} {393} (\bibinfo {year} {1976})}\BibitemShut {NoStop}%
\bibitem [{\citenamefont {Candelas}\ and\ \citenamefont
  {Raine}(1977)}]{Candelas1977}%
  \BibitemOpen
  \bibfield  {author} {\bibinfo {author} {\bibfnamefont {P.}~\bibnamefont
  {Candelas}}\ and\ \bibinfo {author} {\bibfnamefont {D.~J.}\ \bibnamefont
  {Raine}},\ }\href {\doibase 10.1103/PhysRevD.15.1494} {\bibfield  {journal}
  {\bibinfo  {journal} {Phys. Rev. D}\ }\textbf {\bibinfo {volume} {15}},\
  \bibinfo {pages} {1494} (\bibinfo {year} {1977})}\BibitemShut {NoStop}%
\bibitem [{\citenamefont {Davies}\ and\ \citenamefont
  {Fulling}(1977)}]{Davies1977}%
  \BibitemOpen
  \bibfield  {author} {\bibinfo {author} {\bibfnamefont {P.~C.~W.}\
  \bibnamefont {Davies}}\ and\ \bibinfo {author} {\bibfnamefont {S.~A.}\
  \bibnamefont {Fulling}},\ }\href {\doibase 10.1098/rspa.1977.0130} {\bibfield
   {journal} {\bibinfo  {journal} {Proc. Roy. Soc. Lond. A}\ }\textbf {\bibinfo
  {volume} {356}},\ \bibinfo {pages} {237} (\bibinfo {year}
  {1977})}\BibitemShut {NoStop}%
\bibitem [{\citenamefont {Unruh}(1995)}]{Unruh1994}%
  \BibitemOpen
  \bibfield  {author} {\bibinfo {author} {\bibfnamefont {W.~G.}\ \bibnamefont
  {Unruh}},\ }\href {\doibase 10.1103/PhysRevD.51.2827} {\bibfield  {journal}
  {\bibinfo  {journal} {Phys. Rev. D}\ }\textbf {\bibinfo {volume} {51}},\
  \bibinfo {pages} {2827} (\bibinfo {year} {1995})},\ \Eprint
  {http://arxiv.org/abs/gr-qc/9409008} {arXiv:gr-qc/9409008} \BibitemShut
  {NoStop}%
\bibitem [{\citenamefont {Corley}(1997)}]{Corley1996}%
  \BibitemOpen
  \bibfield  {author} {\bibinfo {author} {\bibfnamefont {S.}~\bibnamefont
  {Corley}},\ }\href {\doibase 10.1103/PhysRevD.55.6155} {\bibfield  {journal}
  {\bibinfo  {journal} {Phys. Rev. D}\ }\textbf {\bibinfo {volume} {55}},\
  \bibinfo {pages} {6155} (\bibinfo {year} {1997})}\BibitemShut {NoStop}%
\bibitem [{\citenamefont {Corley}\ and\ \citenamefont
  {Jacobson}(1996)}]{Corley1996b}%
  \BibitemOpen
  \bibfield  {author} {\bibinfo {author} {\bibfnamefont {S.}~\bibnamefont
  {Corley}}\ and\ \bibinfo {author} {\bibfnamefont {T.}~\bibnamefont
  {Jacobson}},\ }\href {\doibase 10.1103/PhysRevD.54.1568} {\bibfield
  {journal} {\bibinfo  {journal} {Phys. Rev. D}\ }\textbf {\bibinfo {volume}
  {54}},\ \bibinfo {pages} {1568} (\bibinfo {year} {1996})},\ \Eprint
  {http://arxiv.org/abs/hep-th/9601073} {arXiv:hep-th/9601073} \BibitemShut
  {NoStop}%
\bibitem [{\citenamefont {Corley}(1998)}]{Corley1997}%
  \BibitemOpen
  \bibfield  {author} {\bibinfo {author} {\bibfnamefont {S.}~\bibnamefont
  {Corley}},\ }\href {\doibase 10.1103/PhysRevD.57.6280} {\bibfield  {journal}
  {\bibinfo  {journal} {Phys. Rev. D}\ }\textbf {\bibinfo {volume} {57}},\
  \bibinfo {pages} {6280} (\bibinfo {year} {1998})},\ \Eprint
  {http://arxiv.org/abs/hep-th/9710075} {arXiv:hep-th/9710075} \BibitemShut
  {NoStop}%
\bibitem [{\citenamefont {Corley}\ and\ \citenamefont
  {Jacobson}(1999)}]{Corley1998}%
  \BibitemOpen
  \bibfield  {author} {\bibinfo {author} {\bibfnamefont {S.}~\bibnamefont
  {Corley}}\ and\ \bibinfo {author} {\bibfnamefont {T.}~\bibnamefont
  {Jacobson}},\ }\href {\doibase 10.1103/PhysRevD.59.124011} {\bibfield
  {journal} {\bibinfo  {journal} {Phys. Rev. D}\ }\textbf {\bibinfo {volume}
  {59}},\ \bibinfo {pages} {124011} (\bibinfo {year} {1999})},\ \Eprint
  {http://arxiv.org/abs/hep-th/9806203} {arXiv:hep-th/9806203} \BibitemShut
  {NoStop}%
\bibitem [{\citenamefont {Unruh}\ and\ \citenamefont
  {Schutzhold}(2005)}]{Unruh2004}%
  \BibitemOpen
  \bibfield  {author} {\bibinfo {author} {\bibfnamefont {W.~G.}\ \bibnamefont
  {Unruh}}\ and\ \bibinfo {author} {\bibfnamefont {R.}~\bibnamefont
  {Schutzhold}},\ }\href {\doibase 10.1103/PhysRevD.71.024028} {\bibfield
  {journal} {\bibinfo  {journal} {Phys. Rev. D}\ }\textbf {\bibinfo {volume}
  {71}},\ \bibinfo {pages} {024028} (\bibinfo {year} {2005})},\ \Eprint
  {http://arxiv.org/abs/gr-qc/0408009} {arXiv:gr-qc/0408009} \BibitemShut
  {NoStop}%
\bibitem [{\citenamefont {Barcelo}\ \emph {et~al.}(2009)\citenamefont
  {Barcelo}, \citenamefont {Garay},\ and\ \citenamefont
  {Jannes}}]{Barcelo2008}%
  \BibitemOpen
  \bibfield  {author} {\bibinfo {author} {\bibfnamefont {C.}~\bibnamefont
  {Barcelo}}, \bibinfo {author} {\bibfnamefont {L.~J.}\ \bibnamefont {Garay}},
  \ and\ \bibinfo {author} {\bibfnamefont {G.}~\bibnamefont {Jannes}},\ }\href
  {\doibase 10.1103/PhysRevD.79.024016} {\bibfield  {journal} {\bibinfo
  {journal} {Phys. Rev. D}\ }\textbf {\bibinfo {volume} {79}},\ \bibinfo
  {pages} {024016} (\bibinfo {year} {2009})},\ \Eprint
  {http://arxiv.org/abs/0807.4147} {arXiv:0807.4147 [gr-qc]} \BibitemShut
  {NoStop}%
\bibitem [{\citenamefont {Finazzi}\ and\ \citenamefont
  {Parentani}(2012)}]{Finazzi2012}%
  \BibitemOpen
  \bibfield  {author} {\bibinfo {author} {\bibfnamefont {S.}~\bibnamefont
  {Finazzi}}\ and\ \bibinfo {author} {\bibfnamefont {R.}~\bibnamefont
  {Parentani}},\ }\href {\doibase 10.1103/PhysRevD.85.124027} {\bibfield
  {journal} {\bibinfo  {journal} {Phys. Rev. D}\ }\textbf {\bibinfo {volume}
  {85}},\ \bibinfo {pages} {124027} (\bibinfo {year} {2012})},\ \Eprint
  {http://arxiv.org/abs/1202.6015} {arXiv:1202.6015 [gr-qc]} \BibitemShut
  {NoStop}%
\bibitem [{\citenamefont {Robertson}\ and\ \citenamefont
  {Parentani}(2015)}]{Robertson2015}%
  \BibitemOpen
  \bibfield  {author} {\bibinfo {author} {\bibfnamefont {S.}~\bibnamefont
  {Robertson}}\ and\ \bibinfo {author} {\bibfnamefont {R.}~\bibnamefont
  {Parentani}},\ }\href {\doibase 10.1103/PhysRevD.92.044043} {\bibfield
  {journal} {\bibinfo  {journal} {Phys. Rev. D}\ }\textbf {\bibinfo {volume}
  {92}},\ \bibinfo {pages} {044043} (\bibinfo {year} {2015})},\ \Eprint
  {http://arxiv.org/abs/1506.02287} {arXiv:1506.02287 [gr-qc]} \BibitemShut
  {NoStop}%
\bibitem [{\citenamefont {Garc\'ia Mart\'in-Caro}\ \emph
  {et~al.}()\citenamefont {Garc\'ia Mart\'in-Caro}, \citenamefont
  {Garc\'ia-Moreno}, \citenamefont {Olmedo},\ and\ \citenamefont
  {Vel\'azquez}}]{prd1}%
  \BibitemOpen
  \bibfield  {author} {\bibinfo {author} {\bibfnamefont {A.}~\bibnamefont
  {Garc\'ia Mart\'in-Caro}}, \bibinfo {author} {\bibfnamefont {G.}~\bibnamefont
  {Garc\'ia-Moreno}}, \bibinfo {author} {\bibfnamefont {J.}~\bibnamefont
  {Olmedo}}, \ and\ \bibinfo {author} {\bibfnamefont {S.}~\bibnamefont
  {Vel\'azquez}},\ }\href@noop {} {}\bibinfo {howpublished} {In
  preparation}\BibitemShut {NoStop}%
\bibitem [{\citenamefont {Chen}\ and\ \citenamefont {Chou}(1997)}]{chen1997}%
  \BibitemOpen
  \bibfield  {author} {\bibinfo {author} {\bibfnamefont {E.}~\bibnamefont
  {Chen}}\ and\ \bibinfo {author} {\bibfnamefont {S.}~\bibnamefont {Chou}},\
  }\href {\doibase 10.1109/22.588606} {\bibfield  {journal} {\bibinfo
  {journal} {IEEE Transactions on Microwave Theory and Techniques}\ }\textbf
  {\bibinfo {volume} {45}},\ \bibinfo {pages} {939} (\bibinfo {year}
  {1997})}\BibitemShut {NoStop}%
\bibitem [{\citenamefont {G\"oppl}\ \emph {et~al.}(2008)\citenamefont
  {G\"oppl}, \citenamefont {Fragner}, \citenamefont {Baur}, \citenamefont
  {Bianchetti}, \citenamefont {Filipp}, \citenamefont {Fink}, \citenamefont
  {Leek}, \citenamefont {Puebla}, \citenamefont {Steffen},\ and\ \citenamefont
  {Wallraff}}]{goppl2008}%
  \BibitemOpen
  \bibfield  {author} {\bibinfo {author} {\bibfnamefont {M.}~\bibnamefont
  {G\"oppl}}, \bibinfo {author} {\bibfnamefont {A.}~\bibnamefont {Fragner}},
  \bibinfo {author} {\bibfnamefont {M.}~\bibnamefont {Baur}}, \bibinfo {author}
  {\bibfnamefont {R.}~\bibnamefont {Bianchetti}}, \bibinfo {author}
  {\bibfnamefont {S.}~\bibnamefont {Filipp}}, \bibinfo {author} {\bibfnamefont
  {J.~M.}\ \bibnamefont {Fink}}, \bibinfo {author} {\bibfnamefont {P.~J.}\
  \bibnamefont {Leek}}, \bibinfo {author} {\bibfnamefont {G.}~\bibnamefont
  {Puebla}}, \bibinfo {author} {\bibfnamefont {L.}~\bibnamefont {Steffen}}, \
  and\ \bibinfo {author} {\bibfnamefont {A.}~\bibnamefont {Wallraff}},\ }\href
  {\doibase 10.1063/1.3010859} {\bibfield  {journal} {\bibinfo  {journal}
  {Journal of Applied Physics}\ }\textbf {\bibinfo {volume} {104}},\ \bibinfo
  {pages} {113904} (\bibinfo {year} {2008})},\ \Eprint
  {http://arxiv.org/abs/https://doi.org/10.1063/1.3010859}
  {https://doi.org/10.1063/1.3010859} \BibitemShut {NoStop}%
\bibitem [{\citenamefont {Parikh}\ \emph {et~al.}(2020)\citenamefont {Parikh},
  \citenamefont {Wilczek},\ and\ \citenamefont {Zahariade}}]{Parikh2020}%
  \BibitemOpen
  \bibfield  {author} {\bibinfo {author} {\bibfnamefont {M.}~\bibnamefont
  {Parikh}}, \bibinfo {author} {\bibfnamefont {F.}~\bibnamefont {Wilczek}}, \
  and\ \bibinfo {author} {\bibfnamefont {G.}~\bibnamefont {Zahariade}},\ }\href
  {\doibase 10.1142/S0218271820420018} {\bibfield  {journal} {\bibinfo
  {journal} {Int. J. Mod. Phys. D}\ }\textbf {\bibinfo {volume} {29}},\
  \bibinfo {pages} {2042001} (\bibinfo {year} {2020})},\ \Eprint
  {http://arxiv.org/abs/2005.07211} {arXiv:2005.07211 [hep-th]} \BibitemShut
  {NoStop}%
\bibitem [{\citenamefont {Parikh}\ \emph {et~al.}(2021)\citenamefont {Parikh},
  \citenamefont {Wilczek},\ and\ \citenamefont {Zahariade}}]{Parikh2020b}%
  \BibitemOpen
  \bibfield  {author} {\bibinfo {author} {\bibfnamefont {M.}~\bibnamefont
  {Parikh}}, \bibinfo {author} {\bibfnamefont {F.}~\bibnamefont {Wilczek}}, \
  and\ \bibinfo {author} {\bibfnamefont {G.}~\bibnamefont {Zahariade}},\ }\href
  {\doibase 10.1103/PhysRevD.104.046021} {\bibfield  {journal} {\bibinfo
  {journal} {Phys. Rev. D}\ }\textbf {\bibinfo {volume} {104}},\ \bibinfo
  {pages} {046021} (\bibinfo {year} {2021})},\ \Eprint
  {http://arxiv.org/abs/2010.08208} {arXiv:2010.08208 [hep-th]} \BibitemShut
  {NoStop}%
\end{thebibliography}%
%------------------------------------------------

%------------------------------------------------
\end{document}